\documentclass[12pt]{article}
\usepackage{titlesec}
\usepackage[T1]{fontenc}
\usepackage[utf8]{inputenc}
\usepackage{booktabs} 
\usepackage{makecell}

    \usepackage{amsfonts}
    \usepackage{amssymb}
    \usepackage{amsbsy}
    \usepackage{amsmath}
    \usepackage{amsthm}
    \usepackage{bbm}
    \usepackage{bm}
    \usepackage{mathtools}
    \newcommand\independent{\protect\mathpalette{\protect\independenT}{\perp}}
\def\independenT#1#2{\mathrel{\rlap{$#1#2$}\mkern2mu{#1#2}}}

    \newtheorem{prop}{Proposition}
    \newtheorem{assum}{Assumption}
\newcounter{subassumption}[assum]
	
	\makeatletter
	\renewcommand{\p@subassumption}{\theassum}
	\makeatother
	
    \newtheorem{cor}{Corollary}

	\newtheorem*{obs*}{Observation}

    \usepackage[flushleft,online,para]{threeparttable}
	\usepackage{graphics}
\usepackage[vlined,linesnumbered,ruled,resetcount]{algorithm2e}
\SetKwInOut{Initialization}{Initialization}

    \usepackage{subfig}
    \graphicspath{{Graphs/}}
    \usepackage{multirow}

    \usepackage{tikz}
    \usetikzlibrary{arrows}
    \usetikzlibrary{decorations.pathreplacing}
    \usepackage{pgfplots}
    \pgfplotsset{compat=1.18}
\usepackage{indentfirst}             
\usepackage{floatrow}
    \usepackage{tabularx}
        \newcolumntype{Z}{>{\centering\arraybackslash}X}
        \newcolumntype{L}{>{\raggedright\arraybackslash}X}
    \usepackage{dcolumn}
        \newcolumntype{d}[1]{D{.}{.}{#1}}
    \usepackage{rotating}                     
    \usepackage{lscape}
    \usepackage{pdflscape}   
\usepackage{authblk}

	\usepackage{csquotes}
	\usepackage[round]{natbib}
	\usepackage{url}
	\let\OLDthebibliography\thebibliography
\renewcommand\thebibliography[1]{
  \OLDthebibliography{#1}
  \setlength{\parskip}{0pt}
  \setlength{\itemsep}{0pt plus 0.3ex}
}

    \usepackage{xcolor}
        \definecolor{darkblue}{rgb}{0,0,0.4}
    \usepackage{hyperref}
        \hypersetup{
            colorlinks = true,
            linkcolor = darkblue,
            citecolor = darkblue,
            pdfborder = 0 0 0,
            pdfdisplaydoctitle = true,
            pdfhighlight = /N,
            pdfpagelayout = OneColumn,
            pdfpagemode = UseNone,
            pdfstartview = {FitH},
            pdfauthor = {{AB, CF \& JR}},
            pdftitle = {{pol-al-cates}},
            pdfsubject = {{}}
        }

    \usepackage{geometry}
    \usepackage{setspace}
    \usepackage[bottom, multiple,flushmargin]{footmisc}  
    \usepackage{verbatim}
    \usepackage[normalem]{ulem}     
    \usepackage{mathpazo}
    \titlespacing*{\section}
{0pt}{0.1\baselineskip}{0.1\baselineskip}
    \titlespacing*{\subsection}
{0pt}{0.1\baselineskip}{0.1\baselineskip}
    \titlespacing*{\subsubsection}
{0pt}{0.1\baselineskip}{0.1\baselineskip}
    \titlespacing*{\paragraph}
{0pt}{0.1\baselineskip}{0.1\baselineskip}
    \usepackage[toc,page]{appendix}
	\usepackage{lipsum}

\DeclareMathOperator*{\argmax}{arg\,max}

\pgfmathdeclarefunction{erf}{1}{%
  \begingroup
    \pgfmathparse{#1 > 0 ? 1 : -1}%
    \edef\sign{\pgfmathresult}%
    \pgfmathparse{abs(#1)}%
    \edef\x{\pgfmathresult}%
    \pgfmathparse{1/(1+0.3275911*\x)}%
    \edef\t{\pgfmathresult}%
    \pgfmathparse{%
      1 - (((((1.061405429*\t -1.453152027)*\t) + 1.421413741)*\t 
      -0.284496736)*\t + 0.254829592)*\t*exp(-(\x*\x))}%
    \edef\y{\pgfmathresult}%
    \pgfmathparse{(\sign)*\y}%
    \pgfmath@smuggleone\pgfmathresult%
  \endgroup
}

\newcommand{\zap}[1]{}

\title{Profit-Aligned CATE Estimation:\\Reconciling Policy Learning and Inference\thanks{We would like to thank Marinho Bertanha, Khai Chiong, Brett Gordon, Ta-Wei Huang, Zhenling Jiang, Joonhwi Joo, Blake McShane, Jason Roos, Navdeep Sahni, Srinivas Tunuguntla, Walter Zhang, Florian Zettelmeyer, Yuting Zhu, and the Kellogg Ad-Tech Lab for helpful comments. Glenn Feng provided outstanding research assistance. Contact: artem.timoshenko@northwestern.edu and caio.waisman@northwestern.edu}}
\author[$\dagger$]{Artem Timoshenko}
\author[$\dagger$]{Caio Waisman}
\affil[$\dagger$]{{\small Kellogg School of Management, Northwestern University}}

\date{\today}
\begin{document}
\maketitle
\vspace{-1cm}
\begin{abstract}
\noindent We propose a framework that aligns Conditional Average Treatment Effect (CATE) estimation with profit maximization. Our method recognizes that, for customers with extreme treatment effects, additional estimation accuracy is unlikely to change the recommended actions. In contrast, accuracy is critical near the decision boundary, where treatment effects are close to treatment costs. Our approach optimizes a novel objective function that concentrates learning capacity along this boundary. The proposed objective is Fisher consistent with respect to the original profit function and yields a consistent estimator for CATEs. Theoretically, our framework unifies standard plug-in optimization and direct policy optimization as limiting cases of the same optimization problem. We further show that entropy-regularized policy optimization is a special case of our framework. This result has a direct practical implication: firms can recover consistent CATE estimates from existing profit-maximization pipelines. We use synthetic data to demonstrate how the proposed framework allows firms to explicitly navigate the trade-off between global prediction accuracy and profit maximization.

\end{abstract}

\bigskip

    
\newpage
\doublespacing
\section{Introduction}\label{sec:introduction}
Firms increasingly personalize marketing actions to maximize incremental profits. Retailers decide which customers should receive discounts, car dealers personalize pricing, and online platforms target sponsored ads \citep{simester2020targeting, dube2023personalized, rafieian2021targeting}. These decisions rely on targeting models that recommend the most effective marketing action for each customer.

There are two conceptually distinct approaches to developing targeting models: ``plug-in optimization'' and ``direct policy optimization.'' The plug-in approach first estimates conditional average treatment effects (CATEs) typically by minimizing a statistical metric, such as the mean squared error (MSE), and then assigns actions based on the largest estimated effect. In contrast, direct policy optimization bypasses CATE estimation entirely; it learns a mapping from customer characteristics to marketing actions by directly maximizing the expected policy value, such as the profit per customer.

Although asymptotically equivalent, these two approaches often yield different targeting policies in finite samples, forcing firms to navigate a difficult tradeoff \citep{flp2022b}. Direct policy optimization is appealing for its explicit focus on profit, but it only provides ``black box'' policy recommendations without interpretable treatment effect estimates. In contrast, plug-in optimization provides interpretable CATEs that offer critical operational advantages. For example, they enable firms to rank-order customers, allowing for seamless adjustments to targeting volume in response to budget cuts or operational constraints without re-estimating the model. These estimates also provide the transparency required to justify the economic logic of algorithmic recommendations to stakeholders \citep{lu2025optimizing}. The challenge is that the plug-in approach inherently sacrifices profit by minimizing estimation error uniformly across the covariate space. This global focus spreads statistical power across all customers, including those whose treatment decisions are already obvious.

To understand this tradeoff, consider a promotional campaign in which the treatment cost is \$5. Clair is a ``clear-cut'' customer with an extreme true treatment effect of \$1,000. Margie is a ``marginal'' customer with an effect of \$6, located just above the decision boundary. The firm needs to estimate Margie's treatment effect very precisely to make the right targeting decision; being off by just a few dollars is enough for a mistake to be made. In turn, the firm can underestimate Clair's treatment effect by hundreds of dollars and still make the right decision. Nevertheless, the global focus of plug-in approaches sacrifices the precision required to correctly classify Margie in order to more accurately estimate Clair's treatment effect, which can lead to suboptimal targeting and reduced expected profits.

In this paper, we propose a framework that reconciles these two optimization approaches, providing firms with a near-optimal targeting policy while simultaneously recovering the magnitude of treatment effects. Our framework recognizes that because direct policy optimization focuses solely on optimal decision-making, it can only identify the sign of the treatment effects net of treatment costs but not their magnitude. To resolve this, we modify the profit maximization problem by replacing the (known) treatment cost with a stochastic threshold. Intuitively, this stochastic distribution forces the estimator to recover the true size of the effect to maximize expected profit across a range of potential costs. This modification yields a smooth surrogate objective that provides statistically consistent CATE estimates while remaining aligned with the firm's true goal of maximizing profit.


Optimization of the surrogate objective induces a form of endogenous weighting we term ``Decision Attention.'' Unlike standard inverse propensity weighting, which depends on covariates to address sampling noise, Decision Attention is endogenous: the weights depend on the model's own predictions. Specifically, the objective function automatically assigns higher weights to marginal customers such as Margie, concentrating learning capacity on the decision boundary. Conversely, it downweights customers with extreme effects like Clair, for whom additional accuracy is unlikely to change the firm's actions, and thus provides lower economic value.

We demonstrate that plug-in optimization and direct policy optimization are not competing alternatives but limiting cases of the same optimization problem. The threshold distribution serves as a ``control knob'' that governs the estimator's focus. When the distribution is fully concentrated at the treatment cost, our optimization problem coincides with direct policy optimization. Conversely, when the distribution is uniform, the problem becomes equivalent to standard plug-in optimization. Rather than forcing firms to choose between paradigms, our method provides a data-driven solution: by treating this distribution as a tunable hyperparameter, firms can use cross-validation to explicitly navigate the tradeoff between global prediction accuracy and local decision quality based on the specific noise and structure of their data.

Practically, this framework reveals that firms do not have to sacrifice profit to achieve transparency. In industry applications, ``pure'' profit optimization is computationally intractable and susceptible to overfitting; instead, practitioners introduce smooth approximations and regularization techniques. We demonstrate that the widely-used entropy-regularized policy optimization with sigmoid smoothing is a special case of our framework \citep{pereyra2017regularizing}. This implies that by simply rescaling the ``black box'' scores that policy learning algorithms already generate, firms can recover consistent CATE estimates without sacrificing policy value. They can maintain their current profit-maximizing pipelines while gaining the rank-ordering flexibility and transparency of causal inference as a byproduct.

Beyond marketing and profit maximization, the proposed framework broadly applies to uplift-based targeting settings with binary treatments. Potential applications include personalized medicine and adaptive clinical trials \citep{dahabreh2016using}, heterogeneous treatment assignment in economics \citep{dehejia2005}, and resource allocation under uncertainty in operations research \citep{zhou2023direct}. Implementation code is available at <link hidden for peer review>.

The remainder of the paper proceeds as follows. First, we review the related literature on estimation of CATEs and policy learning. Section \ref{sec:framework} introduces the targeting problem and our assumptions. Section \ref{sec:method} proposes the surrogate objective and establishes its Fisher consistency. Section \ref{sec:practical_impl} develops the M-estimator, characterizes Decision Attention, and shows that plug-in optimization and direct policy optimization are special cases of our framework. Section \ref{sec:simuls} illustrates the theoretical tradeoffs using synthetic data. Section \ref{sec:conc} concludes.

\subsection*{Related Literature}

CATE estimation and policy learning have largely developed around different objectives: statistical accuracy on one side, profit maximization on the other. Our paper shows these are not competing approaches but limiting cases of a unified framework. We organize the discussion around these two streams.

\subsubsection*{Estimation of CATEs}\label{sec:lit_cates}

The literature on CATE estimation is vast, spanning classical methods such as linear and kernel regressions, matching, and propensity score approaches \citep{imbens2004}, and more recent machine learning methods reviewed by \citet{jacob2021} and \citet{hu2023}. We focus on two aspects of this literature that are most relevant to our paper. First, using the terminology from \cite{ksby2019}, there are generally two types of methods for estimating CATEs. T-learners fit two models to estimate the conditional expectation of the outcome variable on covariates using treatment and control observations separately, while S-learners employ a single model for estimation using all observations concurrently. A complementary line of research develops metalearner architectures, including the X-learner \citep{ksby2019}, the R-learner \citep{nw2021} and the DR-learner \citep{kennedy2024semiparametric}, which improve CATE estimation through cross-fitting and flexible nuisance models. Our proposed method is an S-learner that allows CATE parameterization with flexible functional forms, including neural networks.

Second, most methods rely on minimizing the mean squared error (MSE) or one of its variants created by, for instance, adding a penalization term for regularization. This route is taken, for example, by \cite{ir2013}, \cite{ai2016}, \cite{wager2018estimation}, \cite{nw2021}, and \cite{hmz2023}. In contrast, our estimation criterion is derived directly from the decision maker's true optimization objective, which in our setting is the firm's expected profit.\footnote{In marketing and economics, the insight that MSE can be inadequate for firm decision-making dates back at least to \cite{bg1992}, who studied price-elasticity estimation. The approach of determining the estimation objective based on the decision maker's actual objective has been adopted in several other contexts, including credit scoring \citep{lw2010} and reserve price optimization \citep{clsw2021}.}

Three papers are most closely related to our approach. \cite{eg2022} study linear optimization problems with unknown parameters, such as shortest path and portfolio optimization. They propose a ``smart predict-then-optimize'' framework that integrates the decision maker's payoff optimization with parameter prediction. To ensure computational tractability, \cite{eg2022} introduce surrogate formulations to the original problem and prove their Fisher consistency, which means that the surrogate and the original loss share the same optimal solutions. In a similar spirit, our paper introduces a Fisher consistent surrogate objective for profit maximization that also produces consistent estimates of the CATEs. While \cite{eg2022} focus on linear cost minimization tasks, we study an uplift-based targeting problem and causal estimation, in which the objective is the firm's expected incremental profit.

\cite{ascarza2022eliminating} propose an algorithm for treatment effect estimation that modifies random forests to extract heterogeneity from features unrelated to protected attributes, thus improving fairness in resource allocation. Similarly, our paper adjusts the estimation objective to emphasize the heterogeneity most relevant to the decision context. However, rather than focusing on fairness in the covariate space, we target heterogeneity in the profit outcomes. Our CATE estimator is most accurate near the decision boundary, where customers' treatment effects are close to treatment costs and small estimation errors can change targeting decisions.

Finally, in concurrent research, \cite{can2025} also aim to obtain more accurate CATE estimates around the decision boundary, but focus on the design of the experiment rather than the estimation approach. \cite{can2025} consider a firm that has access to a pool of customers and can sequentially add them to the experiment. The authors propose a sampling procedure that uses customer characteristics to adaptively increase the sample size for CATE estimation via causal forests. In contrast, we take the experimental design and the corresponding propensity scores as given and obtain improved decision-making by altering the estimation procedure.

\subsubsection*{Policy Learning}\label{sec:lit_learning}

The policy learning literature focuses on designing targeting policies that maximize incremental outcomes. A natural approach to policy learning is plug-in optimization, that is, to first estimate CATEs and then use these estimates to define the policy \citep{manski2004, lg2020}. Several studies have derived the resulting properties of such policies; we refer the reader to \cite{hp2020} for a comprehensive survey.

A different route circumvents the estimation of CATEs and instead recovers the optimal policy directly. This is possible by recognizing that the treatment assignment problem is closely related to a weighted classification problem, an insight that has been independently discovered in statistics \citep{zzrk2012}, computer science \citep{sj2015}, and economics \citep{kt2018}. Recent research has extended direct policy learning to incorporate doubly robust estimators \citep{aw2021}, model selection and penalization \citep{mtm2021}, and deep learning \citep{zhang2023}. The distinction also resembles a classic distinction between ``value-function'' and ``policy-value'' optimization in dynamic programming. Our paper is motivated by this literature. We modify the direct policy optimization objective to simultaneously identify the magnitude of treatment effects, yielding consistent CATE estimates while maintaining profit-alignment.

Beyond policy learning, our work connects to the literature on decision-focused learning, which proposes integrating downstream optimization problems into the training of prediction models \citep{donti2017task, wilder2019melding}. These methods share our motivation of aligning estimation with the decision maker's true objective but focus on deterministic combinatorial optimization problems rather than causal inference. Our paper applies this principle to the uplift-based targeting problem, where the decision maker's objective is the firm's expected incremental profit and the unknown parameters are treatment effects rather than cost coefficients. 


\section{The Targeting Problem}\label{sec:framework}

We now introduce the firm's decision problem and the assumptions we maintain throughout the paper. 

\subsection{Problem Formulation}

We consider a firm with a customer base in which each individual customer $i$ is characterized by $k$ observable attributes, $X_i\in\mathbb{X}\subseteq \mathbb{R}^k$. These attributes can include demographic data and past purchasing histories. The firm chooses to assign a binary treatment $W_i\in\{0,1\}$ to each customer $i$, where $W_i=1$ denotes treatment and $W_i=0$ denotes control. This treatment can represent various marketing actions, such as sending a promotional coupon, mailing a catalog, or assigning a personal sales associate.

We denote the potential outcomes with and without treatment by $Y_i(1)$ and $Y_i(0)$, respectively. The potential outcomes can capture different targets for the firm. For example, the outcomes could represent profit per customer, store visits, or customer lifetime value projections (CLV). For exposition, we assume that the potential outcomes represent profit per customer measured in monetary units. We define the conditional average treatment effect (CATE) for customers with characteristics $X_i$ as
\begin{align}\label{eq:cate}
    \tau_0(X_i) \equiv \mathbb{E}\left [Y_i(1) - Y_i(0) \middle \vert X_i \right ]. 
\end{align}
The CATE measures the average incremental gain from treatment for customers sharing characteristics $X_i$. It is the fundamental object of interest in this paper: estimating it accurately and in a way that supports effective targeting decisions is the central challenge we address.

The firm designs a targeting policy, $\pi(\cdot)$, which assigns different treatments to different customers based on observable characteristics, so that $\pi: \mathbb{X}\to\{0,1\}$. Let $P$ denote the joint distribution of $Y_i(1)$, $Y_i(0)$, and $X_i$.  The firm's goal is to find a policy that maximizes expected profit per customer (``policy value''):
\begin{align}\label{eq:dm_prob}
    &\max_{\pi:\mathbb{X}\to\{0,1\}} \mathbb{E}_{P}\left [ \pi(X_i)  \left \{ Y_i(1) - c \right \} + \left \{ 1- \pi(X_i) \right \}  Y_i(0) \right ] \nonumber\\ &\max_{\pi:\mathbb{X}\to\{0,1\}} \mathbb{E}_X\left [ \pi(X_i)  \left \{ \tau_0(X_i) - c \right \}\right]
\end{align}
where $c$ is the known treatment cost per customer (in the treatment condition). The subscript $P$ in the first expectation indicates that it is taken with respect to $Y_i(1)$, $Y_i(0)$, and $X_i$. The second line follows from the law of iterated expectations; it shows that the firm's problem reduces to comparing each customer's CATE to the treatment cost.
We introduce $c$ for expositional reasons. In many applications, this cost can be embedded in the definitions of potential outcomes, which is equivalent to $c=0$ in our formulation.

If the CATEs were known to the firm, the optimal targeting policy, $\pi^*(X_i)$, would assign treatment only to customers with $\tau_0(X_i)\geq c$, that is,
\begin{align}\label{eq:asymptotic_match}
    \pi^*(X_i)=\mathbb{1} \left \{ \tau_0(X_i)\geq c \right \}.
\end{align}
In practice, $\tau_0(X_i)$ is unknown and must be estimated from data. This introduces estimation error and raises the central question of this paper: how should the estimation objective be designed so that errors in the treatment effect estimates translate into as  little profit loss as possible?

\subsection{Data and Assumptions}

We assume that the firm estimates a targeting policy using experimental data. The experiment includes $n$ customers, each randomly assigned to treatment ($W_i=1$) or control ($W_i=0)$ with probability $e_i$:
\begin{align}\label{eq:pscore}
    e_i=e(X_i) \equiv \Pr \left (W_i=1 \middle \vert X_i \right ).
\end{align}
The firm's dataset contains the experimental assignments, $W_i$, customer characteristics, $X_i$, propensity scores, $e_i$, and observed outcomes, $Y_i$, where $Y_i=W_iY_i(1) + (1-W_i) Y_i(0)$. We summarize our notation in Table~\ref{table:notations}.

\begin{table}[h!]
    \centering
    \caption{Notation}
    \renewcommand{\arraystretch}{1.15} 
    \small{
    \begin{tabular}{ c l } 
        \hline\hline
        Notation & Definition \\ 
        \hline
        $n$ & Number of customers in the experiment. \\
        $X_i$ & Covariates for customer $i$, $X_i\in\mathbb{X}\subseteq\mathbb{R}^k$. \\ 
        %
        $W_{i}$ & Treatment assignment for customer $i$ in the experiment (binary).\\
        $Y_{i}(1)\text{, } Y_i(0)$ & Potential outcomes for customer $i$ under treatment and control.\\
        $Y_{i}$ & Observed outcome: $Y_i=W_iY_i(1) + (1-W_i) Y_i(0)$.\\
        $P$ & Joint distribution of $Y_i(1)$, $Y_i(0)$ and $X_i$. \\ 
        $e(X_i)$ & Probability of treatment for customer $i$, $\Pr \left (W_i=1 \middle \vert X_i \right )$. \\ 
        $c$ & Treatment cost per customer in the treatment condition ($W_i=1)$. \\
        $\tau_0(X_i)$ & Conditional average treatment effect (CATE), $\mathbb{E}\left[Y_i(1)-Y_i(0)|X_i\right]$.\\
        $\pi(X_i)$ & Policy recommendation for customer $i$, $\pi:\mathbb{X}\rightarrow\{0,1\} $. \\
         \hline\hline
    \end{tabular}
    \label{table:notations}
    }
\end{table}

Throughout our analysis, we maintain the following assumptions. 

\begin{assum}\label{assum:sutva} Stable Unit Treatment Value Assumption (SUTVA): $Y_i(W_1,\dots,W_i,\dots,W_n)=Y_i(W_i) \textrm{ for all } i$.
\end{assum}

\begin{assum}\label{assum:unconfoundedness}Unconfoundedness:  $Y_i(1),Y_i(0)\independent W_i | X_i \textrm{ for all } i$.
\end{assum}

\begin{assum}\label{assum:overlap}Strong Overlap: $\exists$ $\eta > 0$ such that $\eta < e(X_i) < 1-\eta \textrm{ for all }X_i\in\mathbb{X}$.
\end{assum}


Assumptions~\ref{assum:sutva}--\ref{assum:overlap} are standard in the causal inference literature and are often plausible in marketing experiments. SUTVA rules out interference between customers: the potential outcomes of customer $i$ depend only on $i$'s own treatment assignment, not on those of other customers. This is plausible when promotional treatments are delivered individually (for example, in direct mail or digital channels), and spillovers across customers are negligible. Unconfoundedness and strong overlap are requirements on the experimental design. They ensure that treatment probability is bounded away from 0 and 1 for all $X_i$ and that, conditional on $X_i$, treatment assignments are independent of potential outcomes. We note that for the purposes of obtaining identification of the CATEs, Assumption \ref{assum:overlap} can be relaxed to $e(X)\in(0,1)$ for all $X\in\mathbb{X}$. The stronger version is needed to guarantee the asymptotic properties of the M-estimator developed in Section~\ref{sec:practical_impl}, and it is trivially enforceable in the experiment design.



\subsection{Policy Training and Evaluation}\label{sec:pte}

There are two conceptually different approaches to estimate optimal targeting policies using experimental data. The first approach, plug-in optimization, proceeds in two stages. It first estimates the CATEs, $\widehat{\tau}(\cdot)$, and then assigns treatment to customers whose estimated effect exceeds the treatment cost: $\widehat{\pi}(\cdot)=\mathbbm{1}\{\widehat{\tau}(\cdot)\geq c\}$. The CATE estimates are typically obtained by minimizing the discrepancy between the predicted treatment effects and the observed data. For example, we can parameterize the CATEs as $\tau(X_i;\theta)$, and find $\theta$ by minimizing the following mean squared error (MSE):
\begin{align}\label{eq:mse}
    &\min_{\theta}\frac{1}{n}\sum_{i=1}^n \left [ Y_i^*-\tau(X_i;\theta) \right ]^2
\end{align}
where $Y_i^*$ is a transformed outcome, an unbiased estimate of the individual CATE \citep{ai2016}:
\begin{align}\label{eq:transf_out}
    Y_i^* = Y_i \left (\frac{W_i}{e(X_i)} - \frac{1-W_i}{1-e(X_i)} \right ).
\end{align}

The second approach, direct policy optimization, bypasses CATE estimation entirely. Rather than parameterizing the treatment effect, it directly parameterizes the policy $\pi(X_i;\theta)$ and maximizes a sample analog of the policy value. Using Equation~\eqref{eq:dm_prob}, this objective can be written as:
\begin{align}
    &\max_{\theta}\frac{1}{n}\sum_{i=1}^n\left[ \frac{W_i\pi(X_i;\theta)}{e(X_i)} Y_i + \frac{(1-W_i)[1-\pi(X_i;\theta)]}{1-e(X_i)} Y_i-\pi(X_i;\theta) c\right]\nonumber \\
    &\max_{\theta}\frac{1}{n}\sum_{i=1}^n \left \{  \pi(X_i;\theta)\left [Y_i \left (\frac{W_i}{e(X_i)} - \frac{1-W_i}{1-e(X_i)} \right ) -c \right ]\right \}  \nonumber\\
    &\max_{\theta}\frac{1}{n}\sum_{i=1}^n \left[  \pi(X_i;\theta)\left (Y_i^*-c \right )\right]\label{eq:policy_value_true}
\end{align} 

Intuitively, under Assumptions \ref{assum:sutva}$-$\ref{assum:overlap}, Equation~\eqref{eq:policy_value_true} provides an unbiased estimate of the policy value, as if the policy $\pi(X_i;\theta)$ were evaluated in a field experiment \citep{simester2020efficiently,hmz2023}. Firms can then directly search for the parameters $\theta$ that yield the most profitable targeting policy.

In finite samples, these two approaches often yield substantially different targeting policies \citep{flp2022b}. The bias-variance tradeoff inherent in MSE-based CATE estimation is not aligned with the profit-maximizing objective: the MSE penalizes estimation errors uniformly across the covariate space, whereas profit depends only on the sign of $\tau_0(X_i)-c$ for each customer. Direct policy optimization avoids this misalignment using the ``correct'' objective function.

The theoretical appeal of direct policy optimization comes with a significant practical cost: it produces targeting decisions without CATE estimates. CATE estimates are valuable beyond the binary treatment assignment. First, they provide operational flexibility; for example, firms can easily adjust targeting volumes in response to budget changes or capacity constraints by rank-ordering customers according to the estimated treatment effects. Second, CATEs explain why different customers receive different treatments, providing the transparency needed to justify algorithmic recommendations to managers and regulators \citep{lu2025optimizing}. Third, CATEs can be used to implement operational guardrails, such as fairness or volume constraints, within downstream optimization pipelines. 

We next propose a framework that delivers the profit-alignment of direct policy optimization while simultaneously recovering interpretable CATE estimates, thereby preserving the operational advantages described above.

\section{The Profit-Aligned Surrogate Objective}\label{sec:method}

We now propose a framework that aligns CATE estimation with the firm's profit maximization objective. We first build intuition in a simplified setting without covariates and introduce the surrogate objective. We then extend to heterogeneous customers and present three tractable closed-form specifications.

\subsection{The Surrogate Objective}\label{sec:intuit}

To build intuition, we begin with a simplified setting in which all customers are observationally equivalent  ($X_i=X$ for all $i$), so that the relevant quantity is the scalar average treatment effect (ATE), $\tau_0 \equiv \mathbb{E}[Y_i(1)-Y_i(0)]$. The firm's decision problem reduces to 
\begin{align}
    \max_{\pi\in\{0,1\}} \pi\cdot (\tau_0-c),\label{eq:indiv_standard_profit_payoff}
\end{align}
and the optimal policy is $\pi^*=\mathbbm{1} \{\tau_0 \geq c \}$.  Our goal is to modify this objective so that its unique maximizer is $\tau_0$ itself, recovering the magnitude of the treatment effect rather than just the optimal treatment decision.

We construct the surrogate objective in two steps.

\textbf{Step 1: Change the decision variable.} We recognize that any real-valued estimate $\tau$ implies a targeting policy $\mathbbm{1}\{\tau\geq c\}$. Plugging this decision rule into the profit maximization problem in \eqref{eq:indiv_standard_profit_payoff} and changing the decision variable to $\tau$ yields
\begin{align}
    \max_{\tau\in\mathbb{R}} \mathbbm{1}\left\{\tau\geq c\right\}
    \cdot\left(\tau_0-c\right).\label{eq:indiv_standard_payoff}
\end{align}
This modification alone does not achieve our goal: the set of optimal solutions is
\begin{align}\label{eq:orig_sol}
    \tilde{\tau}=
    \begin{cases}
        [c,+\infty) & \text{if } \tau_0\geq c \\
        (-\infty,c) & \text{otherwise,}
    \end{cases}
\end{align}
so $\tau_0$ is one solution among infinitely many. The objective is flat: any $\tau$ on the correct side of $c$ is equally optimal, regardless of how far it lies from $\tau_0$. The magnitude of the treatment effect is therefore not identified by this problem alone.

\textbf{Step 2: Introduce a stochastic threshold.} We replace the fixed treatment cost $c$ with a stochastic threshold $C\sim F_C(c,\sigma)$. The distribution $F_C(\cdot)$ is from a location-scale family, characterized by the location $c$ and a scale parameter $\sigma$. We then maximize the expectation of the profit objective over $C$:
\begin{align}\max_{\tau\in\mathbb{R}}\;\mathbb{E}_C\left[\mathbbm{1}\left\{\tau\geq C\right\}\cdot\left(\tau_0-C\right)\right] 
    &= \max_{\tau\in\mathbb{R}}
    \int_{-\infty}^{\tau}\left(\tau_0-u\right)
    f_C(u)\,du\label{eq:surrogate_integral}
\end{align}
where $f_C(\cdot)$ is the density of $C$.\footnote{The location-scale assumption is not required for Proposition~\ref{prop:fisher}: any distribution with a density strictly positive near $\tau_0$ suffices. We impose this assumption for convenience, as it yields a single scale parameter $\sigma$ that explicitly governs the estimation-profit tradeoff developed in Section~\ref{sec:practical_impl}.}

Equation~\eqref{eq:surrogate_integral} introduces the surrogate objective that is central to our framework. We next establish its theoretical properties.\footnote{The proposed surrogate objective is inspired by the literature on advertising auctions. Online advertisers often make decisions based on the distribution of competing bids, and for a given bid, the advertiser's outcomes are stochastic rather than deterministic. \cite{wnc2023} leverage this insight to infer the effectiveness of online advertising. Conceptually, our approach requires a targeting policy to ``submit bids'' by choosing $\tau$, and customers are only treated if this ``bid'' exceeds the stochastic threshold $C$, which plays the role of the highest competing bid. The optimal bid coincides with the true ATE, $\tau_0$.}

\begin{prop}[Fisher Consistency]\label{prop:fisher}
Let $f_C(\cdot)$ be strictly positive in a neighborhood of $\tau_0$. Then the surrogate objective 
\begin{equation}
    q(\tau)\equiv\int_{-\infty}^{\tau}
    \left(\tau_0-u\right)f_C(u)\,du
\end{equation}
has a unique maximizer at $\hat{\tau}=\tau_0$. Consequently, the surrogate objective is Fisher consistent with respect to the profit maximization objective in Equation~\eqref{eq:indiv_standard_payoff}.
\end{prop}

\begin{proof}
For any $\tau<\tau_0$, the integrand $(\tau_0-u)$ is strictly positive at $u=\tau$, so $q(\tau)$ is strictly increasing at $\tau$. For any $\tau>\tau_0$, the integrand is strictly negative at $u=\tau$, so $q(\tau)$ is strictly decreasing at $\tau$. Therefore $q(\tau)$ attains its unique maximum at $\hat{\tau}=\tau_0$.\footnote{Fisher consistency is a property of objective functions rather than estimators (Definition~4 of \citealt{eg2022}) and should not be confused with the statistical notion of consistency as convergence in probability (e.g., Definition~6.7 of \citealt{wasserman}). The statistical consistency of our M-estimator is established in Section~\ref{sec:practical_impl}.}
\end{proof}

Intuitively, the standard profit objective ``rewards'' any $\tau$ estimate that falls on the correct side of $c$ equally. For example, if $\tau_0>c$, then every $\tau>c$ earns the same reward $\tau_0-c$, regardless of how far $\tau$ is from $\tau_0$. The estimator therefore receives no signal about the magnitude of the treatment effect from the optimization loss. The stochastic threshold breaks this indifference. If $\tau>\tau_0$, then for realizations $C\in(\tau_0,\tau]$, the integrand $(\tau_0-u)$ is negative, so the expected reward falls below what is achievable at 
$\tau_0$. Similarly, for any $\tau<\tau_0$, the reward can be increased by raising $\tau$. Thus, the unique optimum of the surrogate objective is $\hat{\tau}=\tau_0$, and the magnitude of the treatment effect is identified.

\begin{figure}
    \centering
    \includegraphics[width=0.8\textwidth]{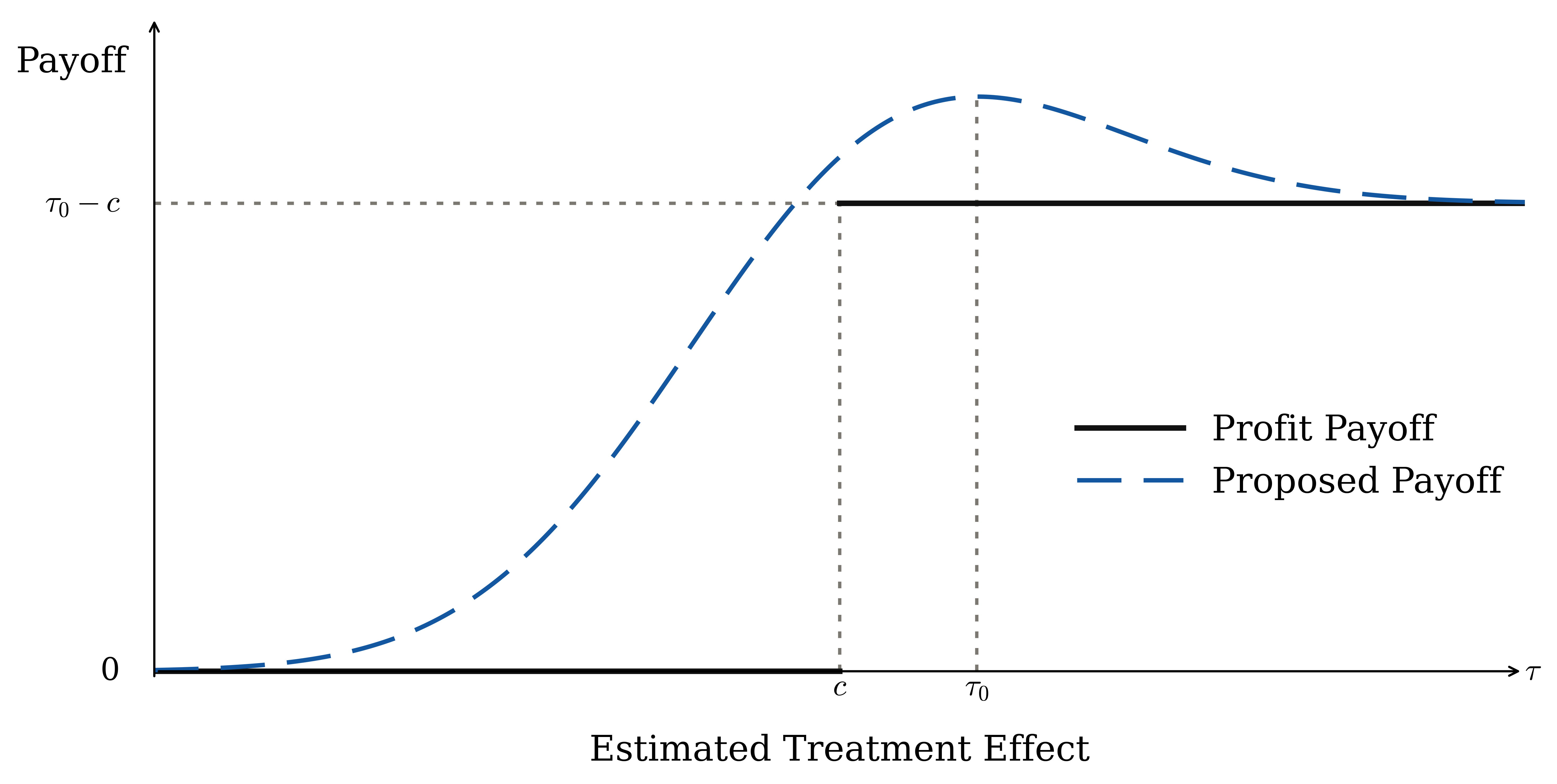}
    \caption{Illustration of the Surrogate Objective Function}
    \label{fig:illus}
\end{figure}

Figure~\ref{fig:illus} illustrates the difference between the standard profit maximization problem and the proposed surrogate. On the horizontal axis, we plot different values of $\tau$, and the vertical axis indicates the value of the optimization objective. The original profit objective (solid line) is a step function: it is flat at zero for $\tau<c$ and flat at $\tau_0-c$ for $\tau\geq c$, without information about the magnitude of $\tau_0$. The surrogate objective (dashed line) is smooth and attains its unique maximum exactly at $\tau_0$, confirming Proposition~\ref{prop:fisher}. The gap between the two curves reflects the cost of smoothing: by using a stochastic threshold $C$ instead of the (known) treatment cost, we introduce a deviation from the true profit objective, but gain both differentiability and point identification of $\tau_0$.\footnote{In Appendix~\ref{app:individual_illustration}, we illustrate how the discrepancy between the standard profit objective and the surrogate function depends on the choice of $F_C(\cdot)$. The scale parameter $\sigma$ controls the width of the gap between the curves.}


\textbf{Remark.} A natural alternative is to combine the profit objective and the MSE directly: $\max_{\tau\in\mathbb{R}} \mathbbm{1}\{\tau \geq c \}(\tau_0-c) - \lambda (\tau_0-\tau)^2$, where $\lambda$ controls the tradeoff between the two. However, this formulation is discontinuous at $\tau = c$, and when extended to treatment effect estimation with covariates, it will exhibit discontinuities throughout the covariate space, making gradient-based optimization infeasible. Our surrogate objective avoids this problem entirely. For any continuously differentiable choice of $F_C(\cdot)$, the objective is smooth and amenable to standard optimization methods.

\subsection{Extension to Heterogeneous Customers} 

To extend the surrogate objective to heterogeneous customers, we take the expectation of Equation~\eqref{eq:surrogate_integral} over the covariate distribution and maximize over functions $\tau:\mathbb{X}\to\mathbb{R}$. Because the expectation $\mathbb{E}_X\left [q\left \{\tau  (X_i) \right \}\right]$ separates across covariate values, this is equivalent to optimizing the surrogate pointwise for each $X_i\in\mathbb{X}$. Proposition~\ref{prop:fisher} then applies at every point, yielding the following result.

\begin{cor}[Fisher Consistency with Covariates]\label{cor:fisher_cov}
Under the conditions of Proposition~\ref{prop:fisher}, the unique population optimizer of the surrogate objective, taken in expectation over $X_i$, is the CATE function $\tau_0(X_i)$ for all $X_i\in\mathbb{X}$.
\end{cor}

These results establish the theoretical foundation of our framework. We next derive closed-form specifications of the surrogate objective and then develop its sample analog as an M-estimator.

\subsection{Closed-Form Representations}\label{sec:specifications}
The surrogate objective in Equation~\eqref{eq:surrogate_integral} admits an equivalent representation that is useful for both interpretation and computation. Integrating by parts, we can rewrite it as
\begin{align}
    \max_{\tau\in\mathbb{R}} F_C(\tau)\cdot\left\{\tau_0-\kappa_C(\tau)\right\},\label{eq:surrogate_kappa}
\end{align}
where $\kappa_C(\tau)\equiv\mathbb{E}[C\mid C\leq\tau]$ is the truncated mean of $C$. This formulation separates the objective into two interpretable components: $F_C(\tau)$ captures the probability of treating the customer, and $\tau_0-\kappa_C(\tau)$ captures the expected net benefit conditional on treatment.

The distribution $F_C(\cdot)$ is a practitioner's choice rather than a feature of the data. For Proposition~\ref{prop:fisher} and Corollary~\ref{cor:fisher_cov} to hold, $f_C(\cdot)$ must be strictly positive in a neighborhood of $\tau_0(X)$ for all $X\in\mathbb{X}$. Since $\tau_0(X)$ is unknown, the simplest way to ensure this is to use a distribution with full support. For computational tractability, we can also choose distributions with closed-form expressions for $F_C(\cdot)$ and $\kappa_C(\cdot)$. We derive these for three distributional families below. Full derivations are provided in Appendix~\ref{app:surrogate_derivations}.

\textbf{Normal.} If $C\sim\mathcal{N}(c,\sigma^2)$, the per-observation contribution to the objective is
\begin{align}\label{eq:ex_normal}
    q(\tau_i) = \Phi(\bar{\tau}_i)(Y_i^*-c)+\sigma\phi(\bar{\tau}_i), \qquad \bar{\tau}_i\equiv\frac{\tau_i-c}{\sigma},
\end{align}
where $\tau_i=\tau(X_i)$, and $\phi(\cdot)$ and $\Phi(\cdot)$ are the probability density function (pdf) and the cumulative distribution function (cdf) of the standard normal distribution, respectively.

\textbf{Logistic.} If $C\sim\mathrm{Logistic}(c,\sigma)$, the contribution is
\begin{align}\label{eq:ex_logis}
    q(\tau_i) = G(\bar{\tau}_i)(Y_i^*-c)+\sigma H \left [G(\bar{\tau}_i) \right ], \qquad \bar{\tau}_i\equiv\frac{\tau_i-c}{\sigma},
\end{align}
where $G(\cdot)$ is the standard logistic cdf and $H(p)=-p\ln p-(1-p)\ln(1-p)$ is the binary entropy function. The logistic distribution is our default choice in all simulations: it satisfies the full support condition and its cdf and truncated mean both have simple closed forms. 

\textbf{Uniform.} If $C\sim\mathcal{U}(\underline{c},\overline{c})$, the contribution simplifies to
\begin{align}\label{eq:ex_unif}
    q(\tau_i) = -\left(Y_i^* - \tau_i\right)^2,
\end{align}
which is invariant to $\underline{c}$ and $\overline{c}$. This case demonstrates that plug-in optimization via MSE is a special case of our framework.

We now turn to estimation, developing the sample analog of these objectives as an M-estimator with standard asymptotic guaranties.

\section{Estimation and Decision Attention}\label{sec:practical_impl}
We now present the sample analog of the surrogate objective as an M-estimator and establish its asymptotic properties. We characterize the estimator's learning mechanism, which we term Decision Attention, and show that plug-in optimization and direct policy optimization arise as special cases of our framework.

\subsection{The M-Estimator}\label{sec:mestimator}
We parametrize the CATE function as $\tau(X_i;\theta)$, where $\theta\in\Theta$ is a parameter vector, and maximize the sample analog of Equation~\eqref{eq:surrogate_kappa}:
\begin{align}\label{eq:estimator}
\widehat{\theta} = \argmax_{\theta\in\Theta} \frac{1}{n}\sum_{i=1}^n \underbrace{F_C\!\left[\tau(X_i;\theta)\right] \Big\{ Y_i^* - \kappa_C\!\left[\tau(X_i;\theta)\right]\Big\}}_{q_i(\theta)}.
\end{align}
For compactness, we write $q_i(\theta)\equiv q \left [\tau(X_i;\theta) \right ]$ to make explicit that the per-observation criterion from Section~\ref{sec:specifications} depends on the data through $\tau(X_i;\theta)$ and $Y_i^*$. Because the objective is a sample average of a per-observation criterion, the proposed estimator is an M-estimator. 

Under standard regularity conditions, the estimator in Equation~\eqref{eq:estimator} is $\sqrt{n}$-consistent and asymptotically normal. We formalize these properties in Propositions~\ref{prop:cons_par} and~\ref{prop:asym_norm_par}, and provide proofs in Appendix~\ref{app:proofs_m_estimators} by verifying the conditions of \citet{nm1994}. 

\begin{prop}[Consistency]\label{prop:cons_par}
    Assume that (i) the data are i.i.d.; (ii) $\Theta$ is compact; (iii) $\mathbb{E}[|Y_i^*|]<\infty$; (iv) $F_C(\cdot)$ is continuous; (v) $\tau(\cdot)$ is continuous in $\theta$; (vi) $\int_{-\infty}^{+\infty}|u|f_C(u)\,du<\infty$; and (vii) $Q(\theta)\equiv\mathbb{E}\left[\frac{1}{n}\sum_{i=1}^n q_i(\theta)\right]$ is uniquely maximized at $\theta_0$. Then $\widehat{\theta}\xrightarrow{p}\theta_0$.
\end{prop}

Proposition~\ref{prop:cons_par} establishes that $\widehat{\theta}$ converges in probability to $\theta_0$, so that the implied CATE estimates $\tau(X_i;\widehat{\theta})$ consistently recover the true $\tau_0(X_i)$ as the sample grows. Proposition~\ref{prop:asym_norm_par} further characterizes the rate of this convergence and the limiting distribution of $\widehat{\theta}$, enabling the construction of confidence intervals for the CATE parameters and standard hypothesis tests on treatment effect heterogeneity.

\begin{prop}[Asymptotic Normality]\label{prop:asym_norm_par}
    Assume the conditions of Proposition~\ref{prop:cons_par} hold. Define $M\equiv\mathbb{E}\!\left[\frac{\partial q_i(\theta_0)}{\partial\theta}\frac{\partial q_i(\theta_0)}{\partial\theta^\top}\right]$ and $B\equiv\mathbb{E}\!\left[\frac{\partial^2 q_i(\theta_0)}{\partial\theta\partial\theta^\top}\right]$. In addition, assume that (i) $\theta_0\in\mathrm{interior}(\Theta)$; (ii) $F_C(\cdot)$ and $\tau(X_i;\theta)$ are $C^2$; (iii) $M<\infty$; (iv) $\left\|\frac{\partial^2 q_i(\theta)}{\partial\theta\partial\theta^\top}\right\|\leq d_B(Y_i^*,X_i)$ with $\mathbb{E}[d_B(Y_i^*,X_i)]<\infty$; and (v) $B$ is nonsingular. Then
\begin{equation}
    \sqrt{n}\left(\widehat{\theta}-\theta_0\right)\xrightarrow{d}N\!\left(0,B^{-1}MB^{-1}\right).
\end{equation}
\end{prop}

The conditions in Propositions~\ref{prop:cons_par} and~\ref{prop:asym_norm_par} are mild. The data conditions, namely i.i.d. observations and a finite first moment for $|Y_i^*|$, are plausible in marketing experiments. The conditions on $F_C(\cdot)$, including continuity, finite moments, and sufficient smoothness for gradient-based optimization, are trivially satisfied because the practitioner chooses this distribution; the normal and logistic families from Section~\ref{sec:specifications} both qualify. Uniqueness of the population maximizer $\theta_0$ holds under correct specification of $\tau(\cdot)$, and is otherwise maintained as an identifying assumption. In Appendix~\ref{app:proofs_m_estimators}, we provide a consistent estimator for the asymptotic variance $B^{-1}MB^{-1}$.

The functional form of $\tau(X_i;\theta)$ is a design choice that governs the flexibility and interpretability of the resulting CATE estimates. The linear specification, $\tau(X;\theta)=X^\top\theta$, is especially appealing when interpretability is important: the estimated coefficients directly quantify how CATEs vary with covariates, which facilitates communication with managers and regulators and supports downstream guardrail constraints. When the true CATE function involves nonlinearities or high-dimensional interactions, neural networks offer greater flexibility and better approximation.\footnote{If the neural network is treated as a concrete parametrization of $\tau(\cdot)$, then Propositions~\ref{prop:cons_par} and 
\ref{prop:asym_norm_par} apply directly. If instead it is treated as a nonparametric approximation, the asymptotic properties follow from sieve M-estimation; see \cite{chen2007} for a review.} In our framework, both specifications can be estimated using standard packages with the custom loss function defined in Equation~\eqref{eq:estimator}.

\subsection{Decision Attention and Implementation}\label{sec:decision_attention}

To characterize how the estimator allocates learning capacity across observations, we derive the gradient of the per-customer criterion $q_i(\theta)$ with respect to $\theta$. For any choice of $F_C(\cdot)$, the gradient takes the form of a weighted residual:
\begin{equation}\label{eq:gradient}
    \nabla_\theta q_i(\theta) = \underbrace{w_i(\theta)}_{\text{weight}} \cdot \underbrace{\bigl(Y_i^* - \tau(X_i;\theta)\bigr)}_{\text{residual}} \cdot \nabla_\theta\tau(X_i;\theta),
\end{equation}
where the scalar weight $w_i(\theta) \equiv f_C\bigl(\tau(X_i;\theta)\bigr)$ is determined endogenously by the model's own predictions. The proposed estimator therefore operates as a weighted least squares problem with endogenous, prediction-dependent weights. We refer to this mechanism as ``Decision Attention.''

For any location-scale distribution $F_C(\cdot)$ with location $c$ and scale $\sigma$, let $f_0(\cdot)$  denote the density of the corresponding standardized distribution. The weight assigned to observation $i$ is $w_i(\theta) \propto f_0\bigl((\tau(X_i;\theta)-c)/\sigma\bigr),$ which depends on how far the predicted treatment effect is from the decision boundary $c$, scaled by $\sigma$. Customers whose predicted effect is close to $c$ receive high weight, concentrating the estimator's capacity where it matters most for targeting decisions. Customers far from $c$ on either side receive low weight, since the firm's action for these customers is unlikely to change regardless of small estimation errors. The parameter $\sigma$ controls the width of the attention region. Large $\sigma$ spreads learning broadly across the covariate space, approaching the equal weighting of OLS; small $\sigma$ concentrates learning tightly around the decision boundary, approaching the binary objective of direct policy optimization.

It is helpful to compare Decision Attention to Inverse Propensity Weighting (IPW). Both methods modify the standard least squares gradient by assigning observation-specific weights to the residual $(Y_i^* - \tau(X_i;\theta))$. In IPW, these weights are pre-specified functions of $X_i$, designed to correct for selection bias by upweighting underrepresented observations. In Decision Attention, the weights are endogenous: they depend on the model's current predictions of $\tau(X_i;\theta)$ and are updated at each gradient step. The two mechanisms also serve different purposes: IPW targets statistical unbiasedness, while Decision Attention trades uniform accuracy across the covariate space for precision near the decision boundary, where targeting mistakes are most costly.

Decision Attention is also related to kernel regression and local polynomial methods, which upweight observations near a focal point. The key distinction is how proximity affects learning. Kernel methods measure distance in the covariate space, assigning a higher weight to observations with $X_i$ close to the focal point. Decision Attention instead assigns a higher weight to observations whose predicted treatment effect $\tau(X_i;\widehat{\theta})$ is close to the decision threshold $c$. Two customers with very different characteristics can receive similar Decision Attention weights if their predicted effects are equally close to the boundary.

The endogenous weighting has two practical implications for implementation. The first concerns overfitting. By concentrating weight on boundary-adjacent customers, Decision Attention effectively reduces the sample size: observations far from $c$ contribute little to the gradient, so the estimator behaves as if trained on a smaller dataset. This reduction in effective sample size increases both the variance near the boundary and the risk of overfitting. Regularization is the natural counterweight. In our applications, we use $\ell_2$ regularization in the linear specification, and incorporate weight decay, dropout, and batch normalization to stabilize training and improve generalizability with neural networks.

Another practical implication concerns initialization. If the model's initial predictions are far from $c$, the weights $w_i(\theta)$ approach zero for most observations, and gradients vanish. We resolve this with a warm start. Specifically, we initialize optimization at a large $\sigma$, which mimics a uniform $F_C(\cdot)$, to obtain a stable initial solution. We then reduce $\sigma$ to gradually shift focus toward the decision boundary. In practice, $\sigma$ is treated as a hyperparameter and selected by cross-validating on held-out policy value. We describe this procedure and illustrate its effect in Section~\ref{sec:simuls}.

\subsection{Special Cases and Connections}\label{sec:bridging}

The M-estimator in Equation~\eqref{eq:estimator} nests the two dominant paradigms in targeting policy design, plug-in optimization and direct policy optimization, as special cases distinguished only by the choice of $F_C(\cdot)$. The scale parameter $\sigma$ controls the balance between treatment effect estimation and profit optimization. Rather than forcing firms to choose between paradigms, our framework allows practitioners to select $\sigma$ via cross-validation to best suit the goals and data structure of their application.

\textbf{Plug-in optimization as a special case.} As shown in Section~\ref{sec:specifications}, when $F_C(\cdot)$ is uniform, the per-observation criterion $q_i(\theta)$ reduces to the negative squared residual $-(Y_i^*-\tau(X_i;\theta))^2$. Maximizing the sample objective is therefore equivalent to minimizing the MSE in Equation~\eqref{eq:mse}, confirming that standard plug-in optimization is a special case of our framework. Intuitively, the uniform distribution assigns equal weight to all potential threshold values, so estimation errors are penalized uniformly across the covariate space, exactly as in OLS.

\textbf{Direct policy optimization as a special case.} As $\sigma\downarrow 0$, the distribution $F_C(\cdot)$ concentrates at the treatment cost $c$, converging to a Dirac delta. In this limit, the surrogate objective collapses to $\mathbbm{1}\{\tau(X_i;\theta)\geq c\}\cdot(\tau_0(X_i)-c)$, which matches the per-customer objective of direct policy optimization. The stochastic threshold $C$ is therefore essential to our framework: without it, the objective does not identify the magnitude of the treatment effect.

The Decision Attention mechanism reveals an important insight about the direct policy optimization problem. As $\sigma$ decreases, the attention weights $w_i(\theta)$ concentrate increasingly on customers near the decision boundary, reducing the effective sample size available for estimation. In the limit, the estimator relies entirely on observations at the boundary, which explains why pure profit optimization is susceptible to overfitting and numerical instability, above and beyond the well-known discontinuity of the profit objective.

\textbf{Connection to entropy-regularized policy optimization.} In practice, firms rarely optimize the pure profit objective directly. The indicator function $\mathbbm{1}\{\tau(X_i;\theta)\geq c\}$ is discontinuous and computationally intractable, so practitioners typically introduce two modifications. First, they replace the indicator with a sigmoid function $G(\bar{\tau}_i)$, which serves as a smooth approximation and corresponds to a softmax layer in neural network implementations. Second, they add an entropy penalty $H\left[G(\bar{\tau}_i)\right]$ to regularize the decision probabilities and prevent overconfident predictions \citep{pereyra2017regularizing}. Together, these two modifications yield exactly the logistic specification of our surrogate objective in Equation~\eqref{eq:ex_logis}. We formalize this connection in the following corollary.

\begin{cor}[Entropy-Regularized Policy Optimization]\label{cor:entropy}
Suppose $F_C(\cdot) = \mathrm{Logistic}(c, \sigma)$. Then the M-estimator in Equation~\eqref{eq:estimator} reduces to maximizing
\begin{align}\label{eq:entropy_obj}
\max_{\theta\in\Theta}\frac{1}{n}\sum_{i=1}^n G(\bar\tau_i)\,(Y_i^* - c) + \sigma\, H \left [ G(\bar\tau_i) \right ],
\end{align}
where $\bar\tau_i = (\tau(X_i;\theta) - c)/\sigma$, $G(\cdot)$ is the standard logistic cdf and $H(p)$ is the binary entropy function. This objective is equivalent to entropy-regularized direct policy optimization with sigmoid smoothing, where $\sigma$ is the regularization weight on the entropy term. 
\end{cor}

This connection has a direct practical implication for firms that already operate entropy-regularized policy learning pipelines. The logit scores $S(X)$ produced by such algorithms are not arbitrary scalars. Our framework implies that they are consistent estimators of the standardized treatment effect, and the CATE can be recovered from the output scores via the simple rescaling
\begin{align}\label{eq:score_interpretation}
\widehat{\tau}(X) = \sigma\cdot S(X) + c.
\end{align}
Firms can therefore obtain consistent CATE estimates by applying Equation~\eqref{eq:score_interpretation} to the estimated scores without changing their existing pipelines.

\section{Illustrative Applications}\label{sec:simuls}
We use synthetic data to demonstrate the theoretical tradeoffs and evaluate the performance of the proposed estimator. We consider two data generating processes (DGPs). The first uses a simple CATE function to illustrate the tradeoff between prediction accuracy and profit optimization under a linear specification. The second features a more complex decision boundary and allows us to compare our method with standard baselines. In both simulations, we evaluate methods on two criteria: incremental \textit{Profit} relative to a blanket no-treatment policy, and \textit{MSE} between the estimated and true CATEs from the DGP.


\subsection{Simulation: Simple CATEs}\label{sec:simulation_quadratic}
\begin{figure}[h!]
    \centering
    \includegraphics[width=0.5\textwidth]{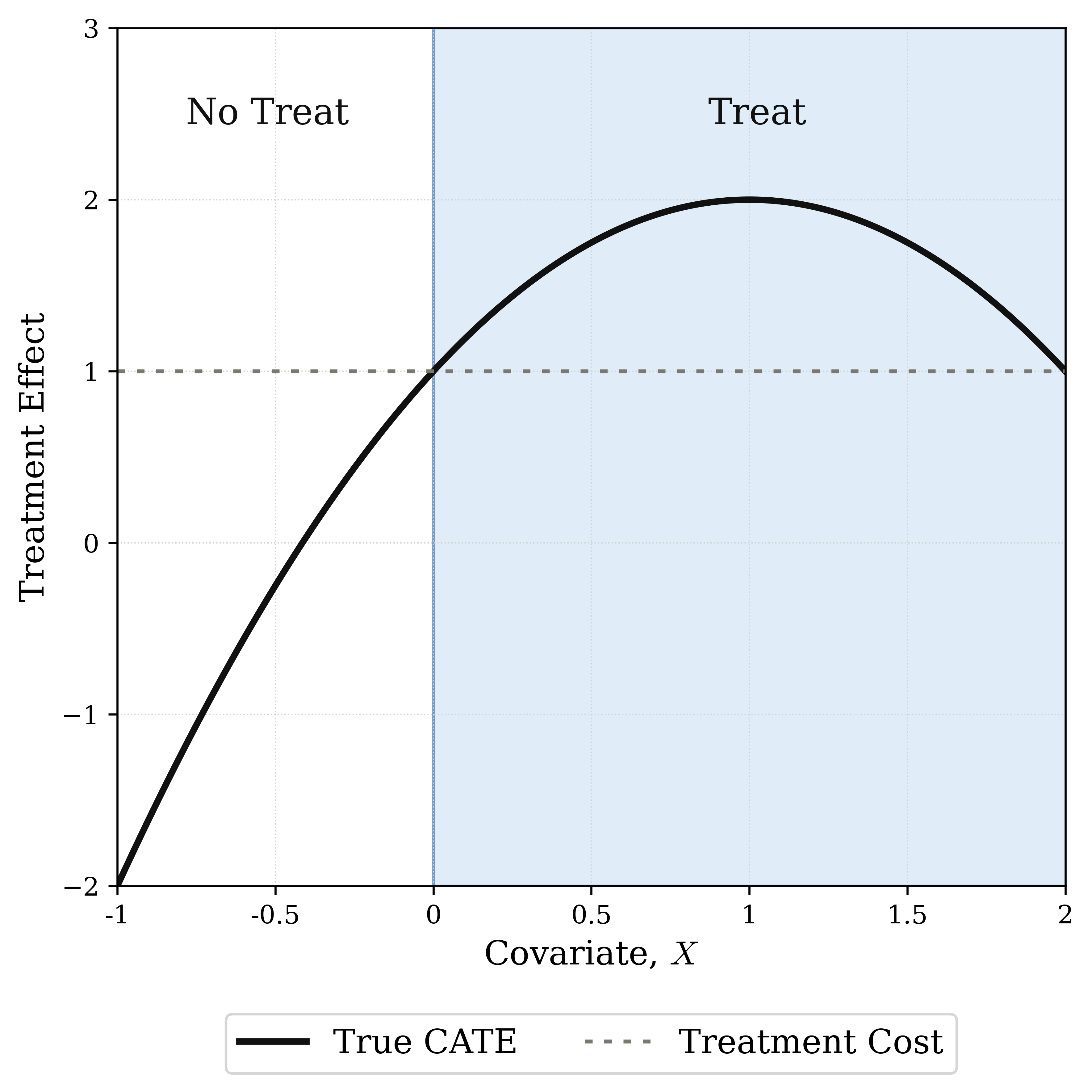}
    \caption{Simple CATEs and the Corresponding Targeting Policy}
    \label{fig:simple_tau}
\end{figure}

We begin with the following DGP:
\begin{align}\label{eq:simple_tau}
    \begin{split}
        Y_i&=1+X_i+ W_i\tau_0(X_i) + \epsilon_i \\
        \tau_0(X_i)&=-X_i^2+2X_i+1 
    \end{split}
\end{align}
where $X_{i}\sim \mathcal{U}(-1,2)$, $W_i\sim\text{Bernoulli}(0.5)$, and $\epsilon_{i}\sim N(0,0.01)$. The treatment cost is $c=1$, which implies that the optimal decision rule is to treat customers if and only if $X\geq0$. Figure~\ref{fig:simple_tau} depicts the true CATEs and the decision boundary.

\begin{figure}[h]
    \centering
    \includegraphics[width=0.5\textwidth]{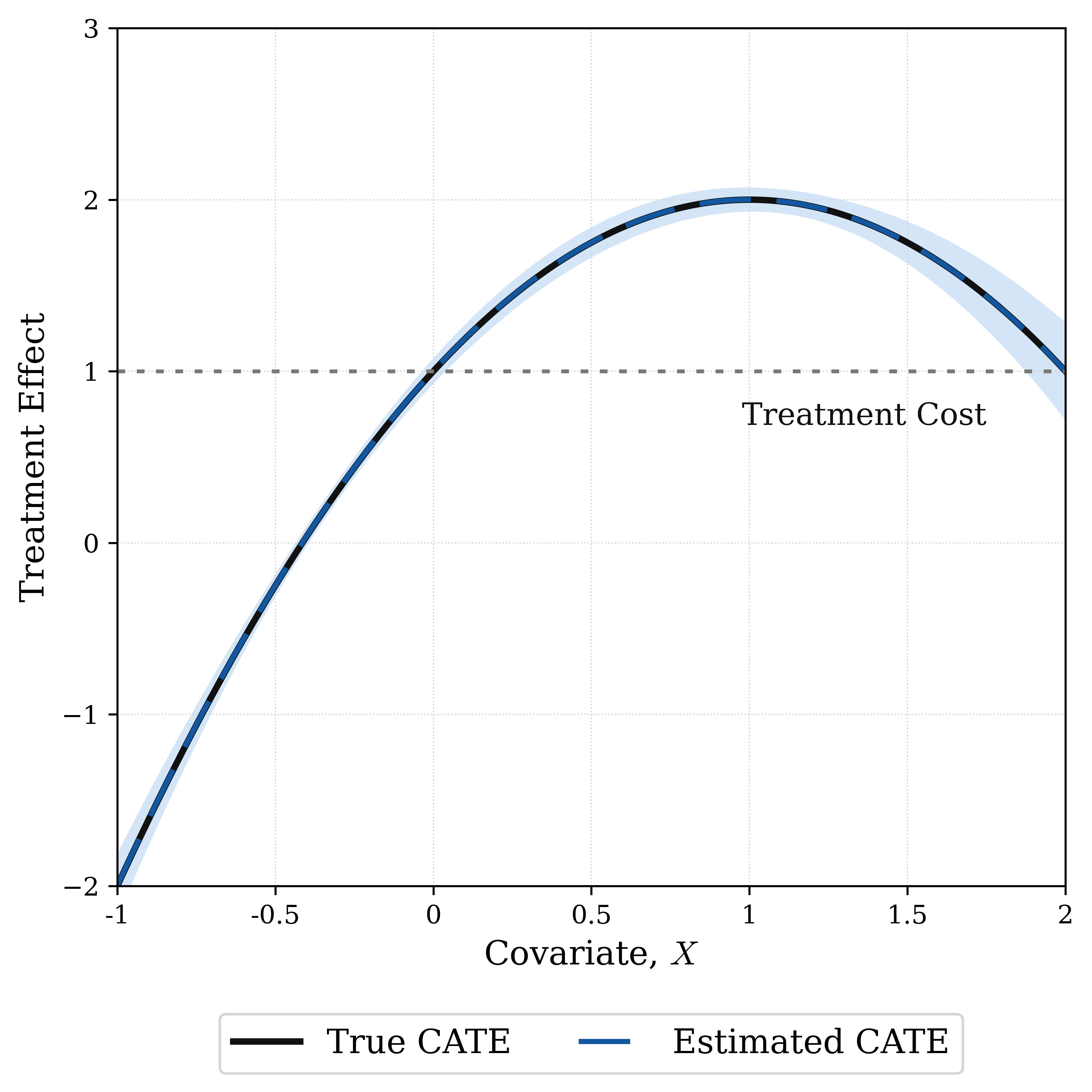}
    \caption{True and Estimated CATEs under Correct Specification}
    \label{fig:simple_approx}
\end{figure}

We first verify that our estimator performs well under correct specification. We estimate the proposed method with a quadratic specification $\tau(X_i;\theta)=\theta_0+\theta_1X_i+\theta_2X_i^2$ and $C \sim \mathrm{Logistic}(c,\sigma)$ with $\sigma=1$ and using 10,000 observations. Figure~\ref{fig:simple_approx} plots the mean estimated CATEs across 1,000 simulation draws together with the true CATEs. The shaded area represents the 95\% confidence interval. Our proposed estimator closely approximates the true CATEs. Consistent with the Decision Attention mechanism, estimation accuracy is highest near the decision boundary ($X=0$), where the estimator concentrates its learning capacity, and variance increases for observations farther from the boundary.

\begin{figure}[h!]
    \centering
    \includegraphics[width=1\textwidth]{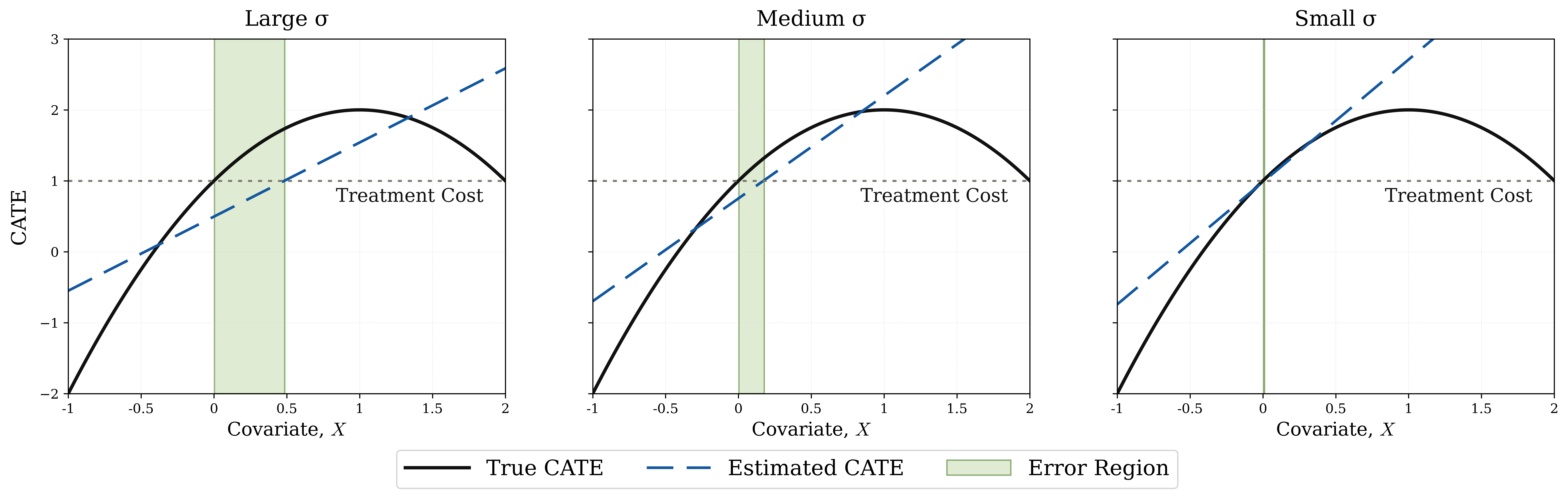}
    \caption{Estimated CATEs with Linear Specification and $F_C(\cdot)=\mathrm{Logistic}(c,\sigma)$}
    \label{fig:simple_vary_sigma}
\end{figure}

Next, we examine the more practically relevant case in which the model is misspecified. We apply our method with a linear specification $\tau(X_i;\theta)=\theta_0+\theta_1X_i$ to the same quadratic DGP. Figure~\ref{fig:simple_vary_sigma} shows how the proposed estimator performs with different values of the parameter $\sigma$. For a single draw of the training data, it plots the true CATEs, the linear approximation obtained from our estimator, and highlights the region where the implied targeting decisions differ from optimal. With large $\sigma$, our method performs similarly to estimating OLS with transformed outcomes, fitting the true CATEs uniformly over the full range of $X$. As $\sigma$ decreases, the estimator concentrates on the decision boundary: the estimated CATEs are more accurate near $X=0$ but less accurate for extreme values of $X$, and the error region shrinks accordingly. With very small $\sigma$, the surrogate objective function approaches a pure profit maximization objective. The error region shrinks, and we observe that the estimated line is almost tangential to the true CATE at $X=0$. 

\begin{figure}[h!]
    \caption{Tradeoff Between MSE and Incremental Profits}
    \centering
    \includegraphics[width=0.6\textwidth]{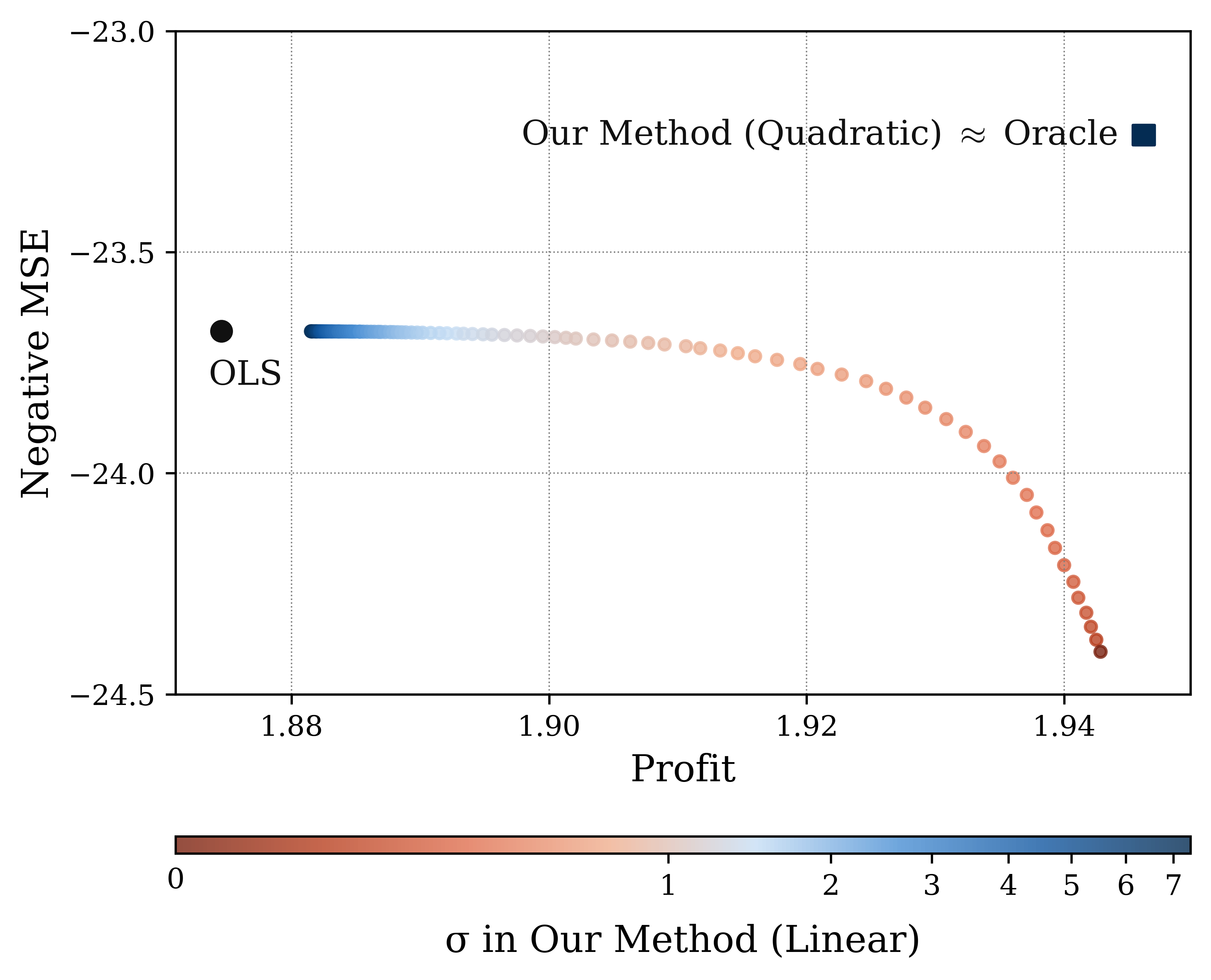}
    \label{fig:simple_frontier}
\end{figure}

Figure~\ref{fig:simple_frontier} further explores the tradeoff between treatment effect estimation and profit optimization. For a single draw of 10,000 observations, we contrast MSE and incremental profit across a grid of $\sigma$ values. Each point corresponds to a different value of $\sigma$, with OLS corresponding to $\sigma=\infty$. The curve traces a ``possibility frontier'' that shows the limit of what can be obtained under the linear specification of $\tau(X;\theta)$. OLS lies at the left end of the frontier, achieving the lowest MSE among linear models at the cost of suboptimal profit; the parameter $\sigma$ controls the balance between CATE estimation and profit optimization, so that linear models can yield higher profit than OLS. The choice of $\sigma$ depends on the firm's goals and data structure and can be determined via cross-validation. When the correct quadratic specification is used, our method approaches the highest performance attainable under this DGP (Oracle), with no tradeoff between the two criteria.

\subsection{Simulation: Complex Decision Boundary}\label{sec:simulation_complex}

We now consider a DGP with nonlinear CATEs and a more complex optimal decision rule:
\begin{align}\label{eq:complex_tau}
    \begin{split}
        Y_i&=W_i\tau_0(X_i) + \epsilon_i \\
        \tau_0(X_i) &= \omega^\top X_i \sin(2.3\cdot\omega^\top X_i)+1.3\\
    \end{split}
\end{align}
where $X_i=(X_{i1},\ldots,X_{i,10})^\top$, $X_{ij}\stackrel{\text{i.i.d.}}{\sim}\mathrm{Uniform}(-1,2)$, and $\omega=\frac{1}{\sqrt{10}}(1,\ldots,1)^\top$. We further assume $W_i\sim\text{Bernoulli}(0.5)$, $\epsilon_{i}\sim N(0,0.1)$, and $c=1$. 

\begin{figure}[h!]
    \centering
    \includegraphics[width=0.65\textwidth]{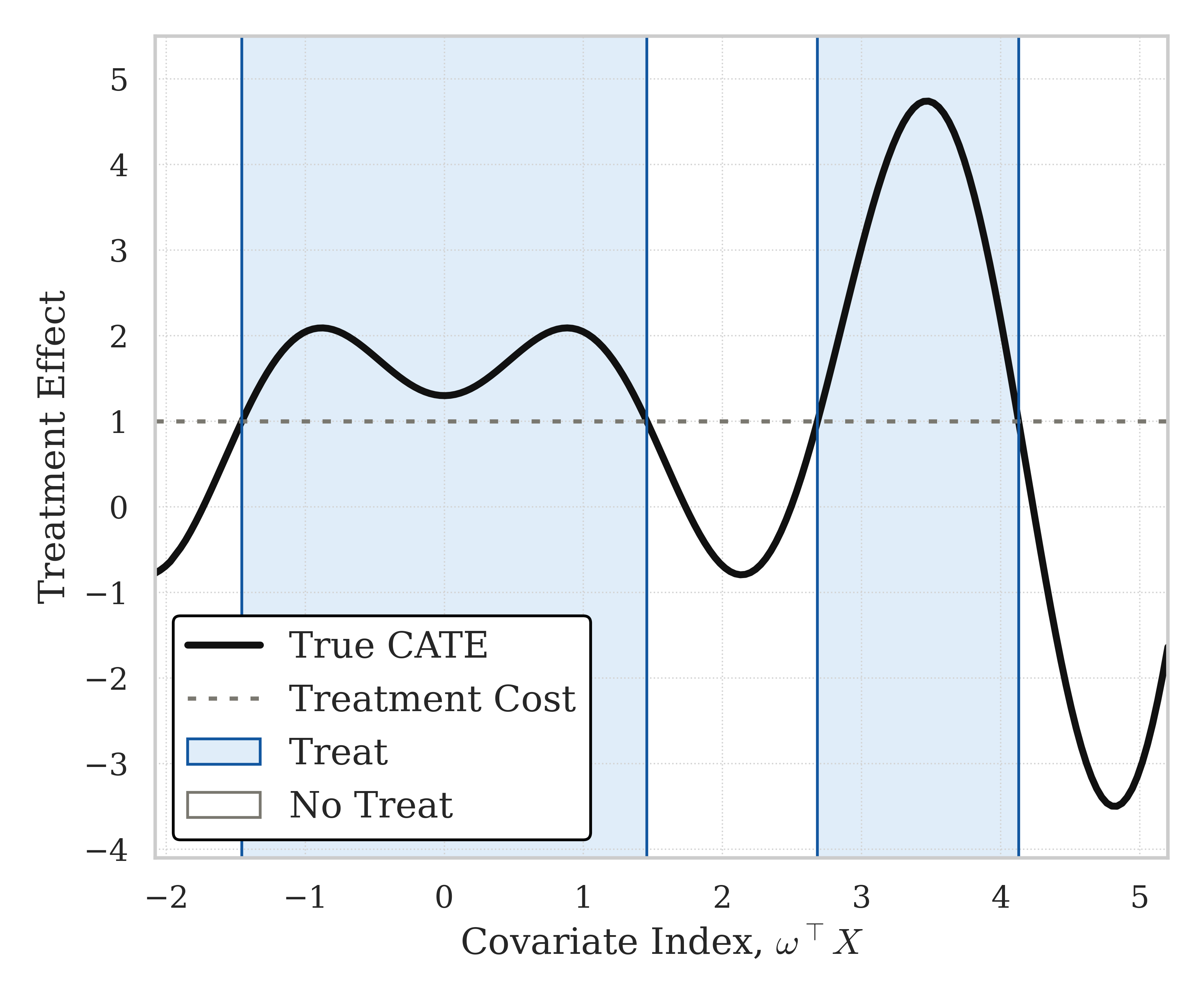}
    \caption{CATE and Complex Optimal Targeting Policy}
    \label{fig:complex_tau}
\end{figure}

Figure~\ref{fig:complex_tau} illustrates the DGP. Compared to Section~\ref{sec:simulation_quadratic}, this specification introduces nonlinearities in the covariates and features no additive separability. The optimal decision rule includes multiple cutoffs in the covariate index $z_i=\omega^\top X_i$, so nonlinear models are expected to outperform linear baselines in both profit and MSE.
 
We compare the performance of the proposed method to three baselines: OLS, gradient-boosting (XGBoost), and causal forest \citep{wager2018estimation}. The OLS and XGBoost are estimated using the transformed outcomes approach, as described in Equation~\eqref{eq:mse}. We calibrate each method using 10,000 observations and then evaluate the estimated treatment effects and targeting policies using the true DGP. 

For our method, we consider linear and neural network parameterizations for CATEs, specify the surrogate distribution $F_C(\cdot)=\mathrm{Logistic}(c,\sigma)$ and select $\sigma$ via cross-validation. The cross-validation grid includes $\sigma=\infty$, which is equivalent to OLS. We report two specifications for the linear case: one with $\sigma$ selected to minimize MSE and one with $\sigma$ selected to maximize profit. For the neural network specification, the choice of $\sigma$ has little effect on profit in this simulation, so we report only the MSE-selected specification. 

\begin{table}[htbp]
\centering
\begin{tabular}{l rr rr}
\toprule
& \multicolumn{2}{c}{\textbf{Profit}}
& \multicolumn{2}{c}{\textbf{MSE}} \\
\cmidrule(lr){2-3}\cmidrule(lr){4-5}
Model & Mean & SD & Mean & SD \\
\midrule
\multicolumn{5}{l}{\textbf{Uniform Baselines}} \\
Mail & -0.004 & 0.001 & -- & -- \\
No Mail & 0.000 & 0.000 & -- & -- \\
\addlinespace
\multicolumn{5}{l}{\textbf{Linear Models}} \\
OLS & 0.262 & 0.030 & 1.496 & 0.004 \\
Our method (linear, MSE) & 0.270 & 0.025 & 1.495 & 0.003 \\
Our method (linear, Profit) & 0.318 & 0.000 & 2.813 & 0.117 \\
\addlinespace
\multicolumn{5}{l}{\textbf{Tree-Based Methods}} \\
Causal Forest & 0.309 & 0.016 & 1.355 & 0.005 \\
XGBoost & 0.283 & 0.006 & 1.184 & 0.017 \\
\addlinespace
\multicolumn{5}{l}{\textbf{Neural Networks}} \\
Our method (NNet) & 0.512 & 0.001 & 0.062 & 0.010 \\
\addlinespace
\multicolumn{5}{l}{\textbf{CATEs from DGP}} \\
Oracle & 0.516 & 0.001 & 0.000 & 0.000 \\
\bottomrule
\end{tabular}
\caption{Method Comparison: Complex Decision Boundary}
\label{tab:complex_methods}
\end{table}

Table~\ref{tab:complex_methods} reports incremental Profit and MSE for each method, averaged across 100 draws of the training data. All targeting methods outperform the uniform baselines. Among linear models, our method with MSE-selected $\sigma$ performs similarly to OLS in both criteria, as expected when cross-validation selects a large $\sigma$. When $\sigma$ is instead selected to maximize profit, our linear method achieves substantially higher profit than OLS, and even outperforms the more flexible tree-based approaches. The tree-based methods achieve lower MSE than any linear specification, but this precision does not translate into superior profit. This result underscores the central insight of our paper: profit maximization requires accurate CATE estimates near the decision boundary, not uniformly across the covariate space.\footnote{In our simulation, tree-based methods can almost perfectly approximate CATEs and yield near-optimal decision rules with one-dimensional covariates, $\dim(X_i)=1$. However, their performance deteriorates when the CATE function involves complex interactions between covariates in multiple dimensions.}

The neural network results directly illustrate the practical implication of Section~\ref{sec:bridging}. Our framework with a logistic $F_C(\cdot)$ is equivalent to entropy-regularized policy optimization with sigmoid smoothing, a combination widely used in practical applications. This equivalence implies that the logit scores produced by such pipelines are consistent estimators of the standardized treatment effects, so that CATE estimates are available as a byproduct of profit maximization at no additional cost. Table~\ref{tab:complex_methods} confirms this: our neural network specification achieves near-Oracle profit while simultaneously recovering low-MSE CATE estimates, a combination that is out of reach for a pure policy optimizer.\footnote{The neural network profit is insensitive to $\sigma$ in this setting. With sufficient training data and model capacity, our approach approximates the true CATE function well and leaves little room for further hyperparameter optimization.}

Together, the simulation results demonstrate how the proposed framework governs the tradeoff between profit alignment and global CATE accuracy. Under a linear specification, selecting $\sigma$ to maximize profit yields meaningful gains over MSE-minimizing estimators, at the cost of reduced global accuracy. The neural network specification shows that with sufficient flexibility, the model can simultaneously attain high profit and low MSE. Together, these results confirm that our framework reconciles profit-aligned targeting with interpretable CATE estimation across model specifications.

\section{Conclusion}\label{sec:conc}
This paper proposes a framework that aligns CATE estimation with profit maximization. By replacing the deterministic treatment cost with a stochastic threshold, we obtain a surrogate objective that is Fisher consistent with respect to the true profit function and produces consistent CATE estimates. The scale of the threshold distribution governs the balance between global prediction accuracy and local profit optimization, so that plug-in optimization and direct policy optimization arise as limiting cases of our framework. When the threshold distribution is logistic, the framework connects to entropy-regularized policy optimization with sigmoid smoothing, and firms can recover consistent CATE estimates from existing pipelines by simply rescaling their output scores.

Simulation results confirm that practitioners can navigate the tradeoff between profit alignment and global CATE accuracy by tuning the threshold distribution. With sufficient model flexibility, high profit and accurate CATE estimation are simultaneously attainable. Although we focus on targeted marketing, the framework applies broadly to any uplift-based targeting setting with binary treatments, including personalized medicine, heterogeneous treatment assignment in economics, and resource allocation in operations research.



\subsection*{Limitations and Future Research}

Our work opens several avenues for future research. First, the current framework focuses on binary treatments. Many practical applications involve more complex decision spaces, such as personalized pricing or multi-channel advertising allocation. Future research could extend the surrogate objective to accommodate multi-valued treatment arms and continuous treatment effects, broadening the applicability of the method beyond binary marketing actions.

Second, our theoretical results establish that the surrogate objective is Fisher consistent with respect to the true profit function and that the resulting M-estimator is $\sqrt{n}$-consistent and asymptotically normal. In finite samples, however, the surrogate may deviate from the true profit maximization goal. Deriving formal bounds on this discrepancy would strengthen the theoretical foundations of the framework and provide practitioners with guidance on when the surrogate objective closely tracks true profit.

Third, our framework is currently formulated for experimental data, where propensity scores are known and the transformed outcomes $Y_i^*$ provide unbiased estimates of the individual CATEs. In observational settings, $Y_i^*$ can be replaced with doubly robust scores \citep{aw2021, chernozhukov2018double}, which remain consistent under either correct specification of the propensity score or of the outcome model. However, incorporating estimated propensity scores introduces an additional source of estimation error, and establishing the asymptotic properties of the resulting estimator requires further theoretical work. A related direction concerns settings where unconfoundedness is not defensible; adapting the framework to incorporate instrumental variables or other identification strategies could broaden its applicability to observational data more generally.

Finally, our method could be integrated with adaptive experimental designs to enhance learning efficiency. Firms could adopt sampling procedures similar to \cite{can2025} to adaptively increase sample sizes in regions of the covariate space that are critical to targeting decisions. Such an integration would complement our estimation approach by concentrating experimental resources where Decision Attention identifies the greatest need for precision.

\newpage
\onehalfspacing
\bibliography{biblio_policy_cates}

\newpage

\appendix
\doublespacing
\renewcommand{\thesubsection}{\Alph{subsection}}
\counterwithin{table}{subsection}   
\counterwithin{figure}{subsection} 
\numberwithin{equation}{subsection}

\section*{Appendix}

\subsection{Additional Illustration of the Surrogate Objective Function}\label{app:individual_illustration}

\begin{figure}[h!]
    \centering
    \includegraphics[width=0.9\textwidth]{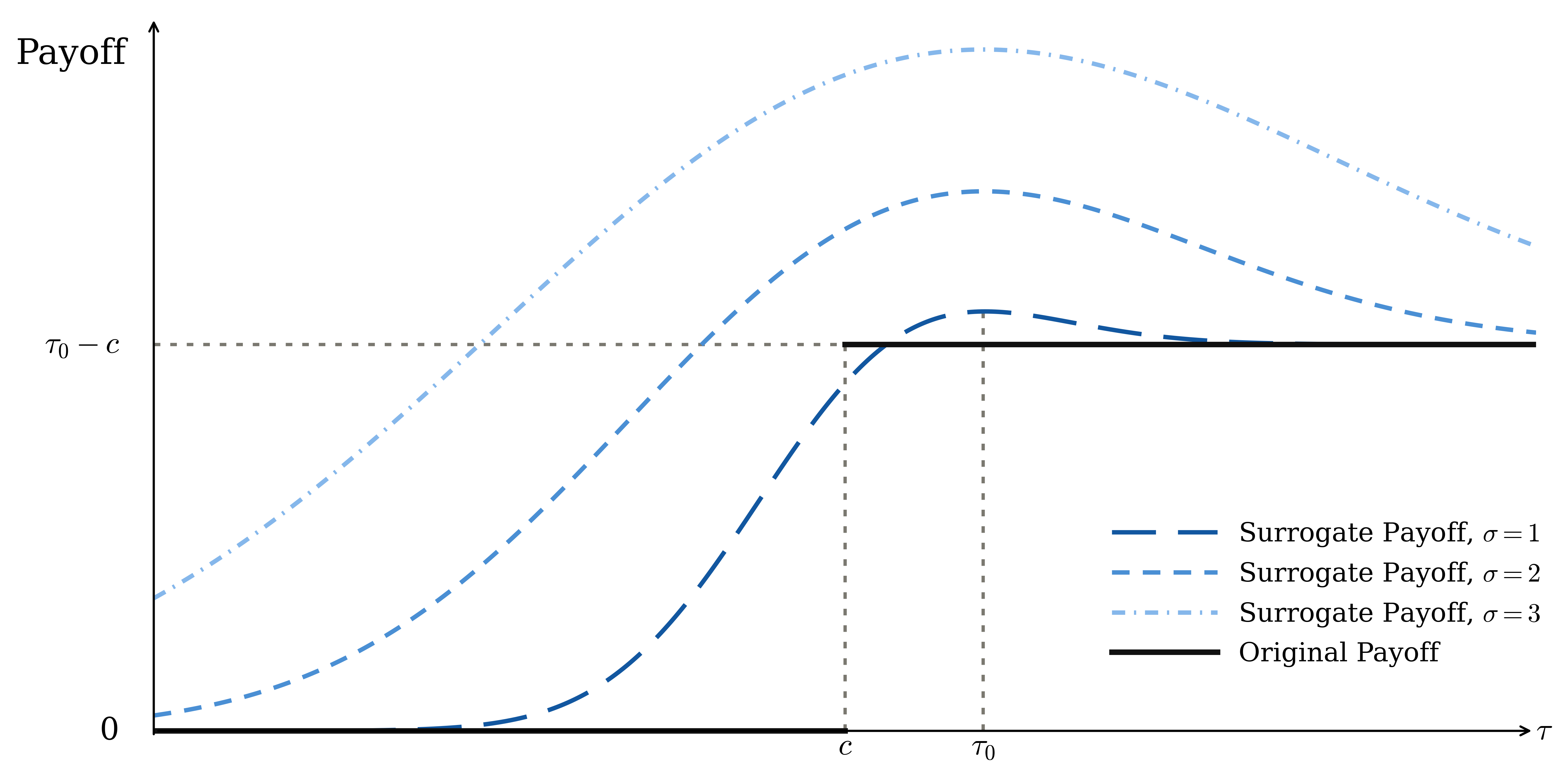}
    \caption{Illustration of the Surrogate Objective Function for Different Values of $\sigma$}
    \label{fig:illus_sigma}
\end{figure}

Figure~\ref{fig:illus_sigma} illustrates the surrogate objective function for $C\sim N(c,\sigma^2)$ and varying values of $\sigma$. The simulation assumes $c=1$ and $\tau_0=2$. The surrogate objective function deviates from the true profit maximization objective more for larger $\sigma$'s. As variance $\sigma$ decreases, the surrogate objective approaches the stepwise function.

%
%
%
%
%

\subsection{Derivations from Section \ref{sec:specifications}}\label{app:surrogate_derivations}

To simplify the notation, we define $\tau_i\equiv\tau(X_i)$, $q_i\equiv q(\tau_i)$, and $\bar\tau_i\equiv\frac{\tau_i-c}{\sigma}$.

\textbf{Normal.} We first rewrite the expression of interest:
\begin{align}\label{eq:term_norm}
   q_i=\int_{-\infty}^{\tau_i}(Y_i^*-u) f_C(u)du&=\int_{-\infty}^{\tau_i}(Y_i^*-c+c-u)f_C(u)du \nonumber \\
   &=(Y_i^*-c)\int_{-\infty}^{\tau_i}f_C(u)du - \sigma \int_{-\infty}^{\tau_i}\frac{u-c}{\sigma}f_C(u)du.
\end{align}

Define $\bar{C}=\frac{C-c}{\sigma}$, so that $\bar{C}$ follows a standard normal distribution. The first integral in (\ref{eq:term_norm}) is
\begin{align}\label{eq:1st_norm}
   \int_{-\infty}^{\tau_i}f_C(u)du=\Pr (C < \tau_i)=\Pr \left (\bar{C} < \bar{\tau}_i \right ) = \Phi\left ( \bar{\tau}_i \right ).
\end{align}

Now consider the second integral:
\begin{align}\label{eq:2nd_norm}
   \int_{-\infty}^{\tau_i}\frac{u-c}{\sigma}f_C(u)du&=\sigma\int_{-\infty}^{\bar{\tau}_i}zf_C(\sigma z+c)dz =\int_{-\infty}^{\bar{\tau}_i}z\phi(z)dz=- \phi\left (\bar{\tau}_i \right ).
\end{align}

Plugging (\ref{eq:1st_norm}) and (\ref{eq:2nd_norm}) back into (\ref{eq:term_norm}) yields Equation (\ref{eq:ex_normal}). 

\textbf{Logistic.} Denote the cdf and pdf of the standard logistic distribution by $G(\cdot)$ and $g(\cdot)$, respectively. Proceeding as in the previous case, we first obtain:
\begin{align}\label{eq:1st_logis}
   q_i=\int_{-\infty}^{\tau_i}(Y_i^*-u)f_C(u)du=G(\bar{\tau}_i)(Y_i^*-c)-\sigma \int_{-\infty}^{\bar{\tau}_i}zg(z)dz
\end{align}

Now notice that:
\begin{align}\label{eq:2nd_logis}
    \int_{-\infty}^{\bar{\tau}_i}zg(z)dz&= zG(z)\Big\vert_{-\infty}^{\bar{\tau}_i}-\int_{-\infty}^{\bar{\tau}_i}G(z)dz \nonumber \\
    &= \bar{\tau}_i G(\bar{\tau}_i)- \ln(1+e^{\bar{\tau}_i}) \nonumber \\
    &= - \bar{\tau}_i \left [ 1 - G(\bar{\tau}_i) \right] + \ln G(\bar{\tau}_i).
\end{align}
Using the logit identity $\bar\tau_i=\ln\frac{G(\bar\tau_i)}{1-G(\bar\tau_i)}$, we can further rewrite this term as follows:
\begin{align}
-\bar{\tau}_i\left[1-G(\bar{\tau}_i)\right]+\ln G(\bar{\tau}_i)
  &= \left[1-G(\bar{\tau}_i)\right]\ln G(\bar{\tau}_i) - \left[1-G(\bar{\tau}_i)\right]\ln(1-G(\bar{\tau}_i))+\ln G(\bar{\tau}_i)\nonumber\\
  &=-G(\bar{\tau}_i)\ln G(\bar{\tau}_i) - \left[1-G(\bar{\tau}_i)\right]\ln(1-G(\bar{\tau}_i))\nonumber\\
  &=H \left [ G(\bar\tau_i) \right]. \label{eq:2nd_logis_entropy}
\end{align}

Plugging~\eqref{eq:2nd_logis_entropy} into~\eqref{eq:1st_logis} yields Equation~\eqref{eq:ex_logis}.

\textbf{Uniform.} Assume that $f_C(\cdot)$ corresponds to the density of a uniform distribution between $\underline{c}$ and $\overline{c}$. Then we have:
\begin{align*}
   q_i&=\int_{-\infty}^{\tau_i}(Y_i^*-u)f_C(u)du \\&=\int_{\underline{c}}^{\tau_i}(Y_i^*-u)\frac{1}{\overline{c}-\underline{c}}du \\
   &=\frac{Y_i^*}{\overline{c}-\underline{c}}\int_{\underline{c}}^{\tau_i}du - \frac{1}{\overline{c}-\underline{c}}\int_{\underline{c}}^{\tau_i}u du \\&=\left(\frac{\tau_i-\underline{c}}{\overline{c}-\underline{c}} \right) Y_i^* - \frac{\tau_i^2-\underline{c}^2}{2(\overline{c}-\underline{c})}\\
   &\propto 2Y_i^*\tau_i-\tau_i^2\\
   &\propto -(Y_i^*-\tau_i)^2,
\end{align*}
where the last two steps recognize that $Y_i^*$ and the bounds $\underline{c}$ and $\overline{c}$ do not depend on the optimization variable $\tau_i$.

\subsection{Derivations from Section \ref{sec:mestimator}}\label{app:proofs_m_estimators}

\subsubsection{Proof of Proposition \ref{prop:cons_par}}

\begin{proof}
    Our proof consists of demonstrating that the conditions of Theorem 2.1 of \cite{nm1994} are met. First, note that continuity of $F_C(\cdot)$ and of $\tau(\cdot)$ in $\theta$ imply that $Q(\cdot)$ is also continuous. 

    Second, it follows that
    \begin{align*}
        |q_i(\theta)|&= \left |F_C \left [\tau(X_i;\theta) \right ]Y_i^* - \int_{-\infty}^{\tau(X_i;\theta)}u f_C(u) du \right | \\
        &\leq F_C \left [\tau(X_i;\theta)\right ]|Y_i^*| +  \int_{-\infty}^{+\infty} |u| f_C(u) du  \\
        &\equiv F_C \left [\tau(X_i;\theta)\right ]|Y_i^*| +  \bar{\mu}_C  \\
        &\leq |Y_i^*|+\bar{\mu}_C \\
        &\equiv d_q(Y_i^*),
    \end{align*}
where $\bar{\mu}_C\equiv\int_{-\infty}^{+\infty}|u| f_C(u) du$. The first inequality follows from $f_C(\cdot)\geq0$ and from the triangle inequality, and the second inequality holds because $F_C \left (\cdot \right ) \leq 1$.
    
Because we have $\mathbb{E} \left [d_q(Y_i^*) \right ]= \mathbb{E} \left [ |Y_i^*| \right ] + \bar{\mu}_C <\infty$, the conditions of Lemma 2.4 of \cite{nm1994} are met. Consequently, it follows that $Q(\theta)$ is continuous and $\sup_{\theta\in\Theta}|\frac{1}{n}\sum_{i=1}^n q_i(\theta)-Q(\theta)|\xrightarrow{p}0$. Therefore, the conditions of Theorem 2.1 of \cite{nm1994} are also met, and $\widehat{\theta}\xrightarrow{p}\theta_0$.
\end{proof}

\subsubsection{Proof of Proposition \ref{prop:asym_norm_par}}

\begin{proof}
    As with Proposition \ref{prop:cons_par}, our proof consists of showing that the conditions for Theorem 3.1 of \cite{nm1994} hold. First, note that the conditions from Proposition \ref{prop:cons_par} plus (i)--(iii) imply that the requirements for the Central Limit Theorem are satisfied, so we obtain $\sqrt{n} \frac{\partial}{\partial \theta}\frac{1}{n}\sum_{i=1}^n q_i(\theta_0) \xrightarrow{d}N \left (0, M  \right )$.

    Next, (iv) implies that Lemma 2.4 of \cite{nm1994} applies again, so that $B(\theta)$ is continuous and $\sup_{\theta \in \Theta} \left | \left | \frac{1}{n}\sum_{i=1}^n \frac{\partial^2 q_i(\theta)}{\partial\theta\partial\theta^\top } - B(\theta) \right | \right | \xrightarrow{p}0$. Therefore, the conditions of Theorem 3.1 of \cite{nm1994} are also satisfied, so it follows that $$\sqrt{n} \left (\widehat{\theta}-\theta_0 \right )\xrightarrow{d}N \left (0, B^{-1} M B^{-1} \right ).$$
\end{proof}

\subsubsection{A Consistent Variance Estimator}
Proposition~\ref{prop:restate_cons_var_par} establishes a consistent estimator for the asymptotic variance.

\begin{prop}[Consistent Estimator for Covariance Matrix]\label{prop:restate_cons_var_par}
    Define $\widehat{B}\equiv \frac{1}{n}\sum_{i=1}^n \frac{\partial^2q_i(\widehat{\theta})}{\partial\theta\partial\theta^\top}$ and $\widehat{M}\equiv \frac{1}{n}\sum_{i=1}^n \frac{\partial q_i(\widehat{\theta})}{\partial\theta} \frac{\partial q_i(\widehat{\theta})}{\partial\theta^\top}$. Assume that the conditions of Propositions \ref{prop:cons_par} and \ref{prop:asym_norm_par} hold. In addition, assume that (i)$\left | \left | \frac{\partial q_i(\theta)}{\partial\theta } \frac{\partial q_i(\theta)}{\partial\theta^\top } \right | \right | \leq d_M (Y_i^*, X_i)$ and (ii) $\mathbb{E} \left[ d_M (Y_i^*, X_i) \right] < \infty $. Then $\widehat{B}^{-1} \widehat{M} \widehat{B}^{-1}\xrightarrow{p}B^{-1} MB^{-1}$.
\end{prop}


\begin{proof}
    The conditions from Propositions \ref{prop:cons_par} and \ref{prop:asym_norm_par} already yield $\widehat{B}^{-1}\xrightarrow{p}B^{-1}$ due to Lemma 2.4 of \cite{nm1994} and the continuous mapping theorem. Conditions (i) and (ii) further imply that we can apply Lemma 2.4 of \cite{nm1994} again, so that $M(\theta)$ is continuous and $\sup_{\theta\in\Theta} \left | \left | \frac{1}{n}\sum_{i=1}^n \frac{\partial q_i(\theta)}{\partial\theta } \frac{\partial q_i(\theta)}{\partial\theta^\top } - M(\theta) \right | \right | \xrightarrow{p}0$, which ensures that $\widehat{M}\xrightarrow{p}M$. Applying the continuous mapping theorem and Slutsky's theorem establishes the result.
\end{proof}

\subsection{Derivatives for Linear Specification}

For the linear specification $\tau(X;\theta)=X^\top\theta$, we can derive closed-form expressions for the derivatives $\frac{\partial q_i(\theta)}{\partial\theta}$ and $\frac{\partial^2q_i(\theta)}{\partial\theta\partial\theta^\top}$ for normal, logistic, and uniform distributions $F_C(\cdot)$. To simplify the exposition, we denote $\tau_i\equiv X_i^T\theta$ and $\bar{\tau}_i\equiv\frac{\tau_i-c}{\sigma}$. 

\textbf{Normal.} From Equation (\ref{eq:ex_normal}), under a normal distribution we now have
\[
q_i(\theta)=\Phi\left(\bar\tau_i\right)(Y^*_i - c)+\sigma\phi\left(\bar\tau_i\right),
\]
where $\Phi(\cdot)$ and $\phi(\cdot)$ correspond to the standard normal cdf and pdf, respectively. The derivatives of $q$ are 
\[
\begin{aligned}
\frac{\partial q_i(\theta)}{\partial\theta}
&= \frac{\phi(\bar\tau_i)}{\sigma}(Y^*_i - \tau_i)X_i,\\
\frac{\partial^2 q_i(\theta)}{\partial\theta\partial\theta^\top}
&= -\frac{\phi(\bar\tau_i)}{\sigma^2}\bar\tau_i(Y^*_i - \tau_i)X_iX_i^\top-\frac{\phi(\bar\tau_i)}{\sigma}X_iX_i^\top.
\end{aligned}
\]

\textbf{Logistic.} From Equation (\ref{eq:ex_logis}) a logistic distribution yields
\[
q_i(\theta)=G(\bar\tau_i)\left(Y^*_i-c\right)+\sigma H(G(\bar\tau_i)),
\]
where $G(u)=1/(1+e^{-u})$ and $g(u)=G^\prime(u)=G(u)\left[1-G(u)\right]$ correspond to the cdf and the pdf of the standard logistic distribution, respectively. The derivatives are
\[
\begin{aligned}
\frac{\partial q_i(\theta)}{\partial\theta} 
&= \frac{g(\bar\tau_i)}{\sigma}(Y^*_i-\tau_i)X_i,\\
\frac{\partial^2 q_i(\theta)}{\partial\theta\partial\theta^\top}
&=-\frac{g(\bar\tau_i)}{\sigma^2}\tanh(\bar\tau_i/2)\left(Y^*_i-\tau_i\right)X_iX_i^\top-\frac{g(\bar\tau_i)}{\sigma}X_iX_i^\top.
\end{aligned}
\]

\textbf{Uniform.} Under a uniform distribution $q_i(\theta)$ collapses to the mean squared loss, so that
\[
q_i(\theta) = -\left[ Y^*_i - X_i^T\theta \right]^2. 
\]
Taking derivatives, we obtain
\[
\frac{\partial q_i(\theta)}{\partial\theta}
=2\left(Y_i^*-X_i^T\theta\right)X_i, 
\quad
\frac{\partial^2 q_i(\theta)}{\partial\theta\partial\theta^\top}
=-2X_iX_i^\top.
\]

\end{document}